%% file: 2-loop.tex
\documentstyle[psfig]{elsart}
\begin{document}
\newcommand{\bare}[1]{\mathaccent"7017{#1}}
\def\i{{\mathrm i}}
\def\c{{\mathrm c}}
%
%
%
\begin{frontmatter}
\title{Minimal renormalization without $\epsilon$-expansion:\\
Amplitude functions for O($n$) symmetric systems\\ in three 
dimensions below~$T_\c$}
\author{Stuart S.~C.~Burnett\thanksref{Canada}},
\author{Martin Str\"{o}sser\thanksref{e-mail-1}},
\author{Volker Dohm\thanksref{e-mail-2}}
\address{Institut f\"{u}r Theoretische Physik, 
Technische Hochschule Aachen,\\
D-52056 Aachen, Germany}
\thanks[Canada]{Present address: Department of Physics,
University of Manitoba, Winnipeg, Manitoba R3T 2N2, Canada; 
e-mail: sburnett@cc.UManitoba.CA}
\thanks[e-mail-1]{e-mail: stroesse@physik.rwth-aachen.de}
\thanks[e-mail-2]{e-mail: vdohm@physik.rwth-aachen.de}
\date{\today}

\begin{abstract}
Massive field theory at fixed dimension $d<4$ is combined with 
the minimal subtraction scheme to calculate the amplitude 
functions of thermodynamic quantities for the O$(n)$ symmetric 
$\phi^{4}$ model below $T_\c$ in two-loop order. 
Goldstone singularities arising at an intermediate stage
in the calculation of O$(n)$ symmetric quantities are shown to 
cancel among themselves leaving a finite result in the limit of 
zero external field. From the free energy
we calculate the amplitude functions in zero field for the 
order parameter, specific heat and helicity modulus 
(superfluid density) in three dimensions.
We also calculate the $q^2$ part of the
inverse of the wavenumber-dependent transverse susceptibility 
$\chi_{{}_T}(q)$ which provides an independent check of our 
result for the helicity modulus. The two-loop contributions to 
the superfluid density and specific heat below $T_\c$ turn
out to be comparable in magnitude to the one-loop contributions, 
indicating the necessity of higher-order calculations and 
Pad\'e-Borel type resummations.\\[5mm]
\end{abstract}
\end{frontmatter}

\input{intro}

%
\input{FT_3d}

%
\input{free}
%
\input{M0}
%
\input{C}
%
\input{rho_s}

%
\input{discussion}
%
\input{appA}

%
\input{appB}
%
\input{appC}
%
\input{appD}

%
\input{refs}
%
\end{document}

%% file: intro.tex
%
%
\section{Introduction}
\label{sec:intro}
Field-theoretical calculations in the context of critical phenomena 
\cite{zinn-justin89,amit84} 
fall into
two categories (i) those based on expansions about the upper 
\cite{wilson72} or lower \cite{brezin76}
critical dimension ($\epsilon$-expansions) and (ii) those that
are carried out at fixed dimension \cite{parisi73}. Within these 
two approaches,
one can further distinguish between two types of renormalization: 
the use of renormalization conditions \cite{brezin76'} and the 
minimal subtraction scheme \cite{thooft72}.
Whereas the
$\epsilon$-expansion has been used extensively with both
types of renormalization, the theory at fixed dimension is usually 
presented within the framework of renormalization conditions 
\cite{zinn-justin89,parisi73,baker76,guillon77,bagnuls85,bagnuls87,muenster94,guida96}.

In this paper we shall use an approach 
\cite{dohm85,dohm85',schloms89,schloms90} that combines
the theory at fixed dimension with the minimal subtraction scheme.
While any renormalization scheme implies a (scheme-dependent)
decomposition of correlation functions
into exponential parts and amplitude functions our present approach 
is especially advantageous
since here the exponential parts
are determined entirely from pure dimensional poles which remain 
unchanged in extensions of the theory
from $T>T_\c$ to $T<T_\c$ as well as to finite $k$, $\omega$ and 
$L$ (wavenumber, frequency and system size) 
\cite{dohm85,dohm85',schloms89,schloms90,dedominicis78,dohm87,pankert89,dohm93}.
Accordingly, this concept of a minimally
renormalized theory at fixed dimension has been applied successfully
not only to static critical phenomena 
\cite{krause90,halfkann92,haussmann92}
but also to dynamics in equilibrium 
\cite{dohm85',pankert89,dohm91,taeuber92,moser} and 
nonequilibrium \cite{haussmann91},
finite-size effects \cite{esser95,chen_to_be,koch96} and surface 
critical phenomena \cite{frank91,mohr96}.
Here, we use this approach to calculate static amplitude functions 
of O$(n)$
symmetric systems below $T_{C}$. For $n>1$ these systems are of
particular interest because of the massless Goldstone modes 
\cite{goldstone61} governing the long-distance properties.

The O($n$) symmetric $\phi^{4}$ model \cite{zinn-justin89,amit84} 
describes the most common examples of critical behavior:
in liquid-gas systems ($n=1$), superfluid $^{4}$He ($n=2$) 
and isotropic magnets ($n=3$). Since
the critical properties of $^{4}$He near the $\lambda$-transition 
can be measured with high accuracy not only near a single critical 
point $T_\lambda$ \cite{lipa96} but also near a whole 
(pressure-dependent) line $T_\lambda(P)$ of
critical points \cite{ahlers78} the opportunity exists for 
the universality 
predictions of the renormalization group (RG) theory to be tested 
at a highly quantitative level.
Owing to the wide range of applicability of the RG concept 
\cite{zinn-justin89} this test is of relevance
not only to statistical physics but equally well to elementary 
particle physics
and condensed matter physics.
In an effort to obtain the most accurate data possible, 
high-precision
measurements are to be carried out in a
microgravity environment \cite{lipa95}. The corresponding 
theoretical challenge is to
calculate as accurately as possible the
properties of the $\phi^{4}$ model appropriate for a comparison 
with the helium data. This includes not only the well-known 
exponent 
functions \cite{zinn-justin89,amit84,schloms89} whose fixed points
values determine the critical exponents
but also the less well-known amplitude functions 
\cite{dohm85,dohm85',schloms89,schloms90,krause90,halfkann92} which 
determine the equation of state and which contain information about
asymptotic ratios of leading and subleading amplitudes 
\cite{privman91}.
They are also needed for a description of nonasymptotic critical
properties. The latter are important in order to distinguish 
universal from non-universal
properties of critical behavior over a wide temperature range 
\cite{dohm87,schloms87}.

Accurate RG calculations of amplitude functions for both $T>T_\c$ 
and $T<T_\c$ have been
performed within the $d=3$ theory only for the case 
$n=1$
\cite{bagnuls85,bagnuls87,muenster94,guida96,krause90,halfkann92}.
In this case Borel resummations yielded accurate results
because the perturbation series are known to high enough order 
(five loops).
For $n>1$, only amplitude functions \em above\/ \em
$T_{C}$ have been obtained with comparable accuracy 
\cite{bagnuls85,krause90};
{\em below} $T_{C}$, the complications due to Goldstone 
singularities have prevented calculations
to sufficiently high order for resummation methods to
yield accurate results. For example, the amplitude functions for 
the ($n>1$) 
order parameter and superfluid density have been computed in the
$d=3$ theory only in one-loop order \cite{schloms90,schloms87} 
where Goldstone singularities do not yet arise.

Our aime here is to present the first step towards filling this 
gap of
theoretical knowledge for $n>1$ below $T_{C}$ for the O($n$) 
symmetric 
$\phi^{4}$ model in three dimensions. Specifically, we calculate 
in two-loop order
the amplitude functions of the order parameter, the specific 
heat and the helicity modulus introduced by Fisher, Barber and 
Jasnow \cite{fisher73}.
We also calculate the $q^2$ part of 
$\mathaccent"7017{\chi}_{{}_T}(q)^{-1}$ 
where $\mathaccent"7017{\chi}_{{}_T}(q)$ is the wavenumber 
dependent transverse 
susceptibility. This quantity enters Josephson's definition 
\cite{josephson66} 
of the superfluid density and 
provides an independent check of our calculation of the helicity 
modulus.

These calculations serve three purposes. First, by comparing 
the zero-, one- and two-loop terms
we can study the reliability of the low-order perturbation theory.
Whereas the two-loop contribution to the $q^2$ part of 
$\mathaccent"7017{\chi}_{{}_T}(q)^{-1}$ turns out to be
very small we find that the two-loop contributions to the 
superfluid density and specific
heat below $T_\c$ are comparable in magnitude to the one-loop 
contributions, thus
indicating the necessity of higer-order calculations and 
Pad\'e-Borel type resummations.
Second, we can anticipate some of the technical difficulties that 
will also appear
in higher-order calculations, where the amount of computational
labor required is substantially greater. These difficulties 
include the 
removal of ultraviolet divergences in three dimensions and 
of spurious infrared (Goldstone) divergences, both of which first
appear here at the two-loop level. Third, we provide part of the 
information
for a future study of the $n$-dependence of amplitude functions 
below $T_\c$ beyond
one-loop order to be carried out in a combined analysis of known 
Borel-summed
amplitude functions for $n=1$ and of low-order contributions for 
$n>1$.

This paper is organized as follows. 
In Sec.~\ref{sec:FT_3d}, we summarize some relevant aspects
of the theory at fixed dimension.
In Sec.~\ref{sec:free_2}, because of our interest in computing
the helicity modulus (superfluid density), we study the
free energy for a state with a twisted order parameter.
In Sec.~\ref{sec:M0}, we use the free energy
to derive the bare two-loop expression for the equation of state 
and the longitudinal 
susceptibility. We follow this, in Secs.~\ref{sec:C} and 
\ref{sec:rho_s}, by a 
calculation of the
specific heat, of the helicity modulus and
of the $q^2$ part of $\mathaccent"7017{\chi}_{{}_T}(q)^{-1}$. 
We find several 
contributions (diagrams)
that have Goldstone divergencies which we regulate by
use of an external field. For rotationally invariant quantities 
we show that these divergencies
cancel leaving a finite result in the limit of zero field.
For the order parameter, specific heat,
helicity modulus (superfluid density) and for the $q^2$ part of 
$\mathaccent"7017{\chi}_{{}_T}(q)^{-1}$, 
we determine their respective amplitude functions up
to two-loop order. The results and conclusions are presented in 
Sec.~{\ref{sec:discussion}
and details of the calculations are given in the appendices. 
Further details are given in 
Ref.~\cite{stroesser96}.
A short summary of some of the results of the present paper 
has been presented in Ref.~\cite{LT21}.

%% file: FT_3d.tex
%
%
\section{Field theory in three dimensions}
\label{sec:FT_3d}
In this section, we outline the strategy of our calculations within
the minimally
renormalized massive field theory at fixed $2<d<4$.
Consider the O$(n)$ symmetric $\phi^{4}$ model in the presence of 
an external
field $\vec{h}_0({\bf x})$, as described by the standard
Landau-Ginzburg-Wilson functional \cite{zinn-justin89,amit84}
\begin{equation}
{\cal H}\{\vec\phi_0({\bf x})\} = \int_V \mbox{d}^{d}x
\left(\frac{1}{2}r_0\phi_0^2 + \frac{1}{2}\sum_{i}
(\nabla\phi_{0i})^2
+u_0(\phi_0^2)^2 -\vec{h}_0\cdot\vec\phi_0\right), \label{eq:LGW}
\end{equation}
where 
$\vec\phi_0({\bf x})=(\phi_{01}({\bf x}),\ldots,\phi_{0n}({\bf x}))$
is an $n$ component vector subject to the statistical
weight $\sim\exp(-{\cal H})$ and $V$ is the volume. The spatial 
variations of 
$\vec\phi_0({\bf x})$ are restricted to wavenumbers less than some 
cutoff 
$\Lambda$. The Gibbs free energy per unit volume 
(divided by $k_B T$) is
\begin{equation}
F_0(r_0,u_0,\{\vec{h}_0({\bf x})\}) = -V^{-1} \ln\int\! {\cal D}
\vec{\phi_0}({\bf x})\,\exp(-{\cal H}\{\vec\phi_0({\bf x})\})
\label{eq:gibbs}
\end{equation}
and is related to the Helmholtz free energy $\Gamma_0$ 
per unit volume via
\begin{eqnarray}
\Gamma_0(r_0,u_0,\{\langle\vec\phi_0({\bf x})\rangle\}) &=&
F_0(r_0,u_0,\{\vec{h}_0({\bf x})\})
+V^{-1} \int_V\! \mbox{d}^{d}x \, \vec{h}_0({\bf x})\cdot
\langle\vec\phi_0({\bf x})\rangle \, , \label{eq:helmholtz} \\
\langle\vec\phi_0({\bf x})\rangle &=& \mbox{}
-V\, \frac{\delta F_0}{\delta \vec{h}_0({\bf x}) } \,. 
\label{eq:h2phi}
\end{eqnarray}
$\Gamma_0$ is the generating functional for vertex functions and can
be obtained perturbatively from the negative sum of all 
one-particle irreducible (1PI) vacuum
diagrams. We shall consider always the bulk limit 
$V\rightarrow\infty$. We shall further suppose that finite-cutoff 
effects are negligible (although this may not be justified in 
certain cases \cite{dohm93,bagnuls94,anisimov95})
and that all integrals are evaluated in the
limit $\Lambda\rightarrow\infty$ according to the prescriptions of 
dimensional regularization  \cite{zinn-justin89,amit84}.
The ultraviolet divergences are thus manifested as poles in the 
dimension~$d$ at discrete 
values $d_{l}=4-2/l$ (where $l=2,3,\ldots)$ if the vertex functions 
are considered as
functions of $r_0$  \cite{bagnuls85,bagnuls87,schloms89,symanzik73}.
Our ultimate goal is to calculate the vertex functions at $d=3$ as 
functions of the correlation
length as this will ensure
that the perturbative expansions have no poles and are (presumably) 
Borel resummable  \cite{bagnuls85,bagnuls87,schloms89,schloms90}.
Our strategy will be to treat the $d=3$ poles at the level 
of the bare free energy and use the resulting finite
expression to derive bare expressions for the other quantities 
directly in three dimensions.

One way to remove the poles
at $d_l<4$ is to rewrite the perturbation
series in terms of the variable $r_0-r_{0\c}$ where $r_{0\c}$ is 
the critical value of $r_0=r_{0\c}+a_0t$ with $t=(T-T_\c)/T_\c$ 
being the reduced temperature.
The structure of $r_{0\c}$
is \cite{symanzik73} $r_{0\c} = u_0^{2/\epsilon}S(\epsilon)$,
where $S(\epsilon)$ is a dimensionless function which is finite for 
$\epsilon=4-d>0$ except for the poles at $d=d_l$ $(\epsilon=2/l)$ 
which cancel the poles mentioned above.
The role played by $r_{0\c}$ in the $d=3$ theory was discussed in 
detail by Bagnuls and Bervillier \cite{bagnuls85}, by Bagnuls et 
al.~\cite{bagnuls87} and, in the context of the minimal
renormalization, by Schloms and Dohm  \cite{schloms89}.
Another possibility, which we shall use below, is to use a shifted 
variable $r_0'$
(given in Eq.~(\ref{eq:r0_pr}) below) which differs from 
$r_0-r_{0\c}$ only by a conveniently 
chosen finite constant \cite{bagnuls87,halfkann92}.
Expressed in terms of the variables
$r_0-r_{0\c}$ or $r_0'$, the unrenormalized vertex functions are 
finite for $d<4$ but still have poles at 
$d=4$ which can be absorbed by the $Z$-factors either by use of 
($d=3$) renormalization
condititions \cite{bagnuls85,bagnuls87} or within the minimal 
subtraction scheme
\cite{dohm85,dohm85',schloms89}. Although in the latter approach 
the $Z$ factors are
determined by ($d=4$) pole terms $\sim \epsilon^{-n}$ this does 
not imply the
necessity of an $\epsilon$ expansion, as shown in 
Ref.~\cite{schloms89}.
The $Z$-factors connect the bare and the renormalized quantities 
in the usual way
\begin{equation}
r = Z_r^{-1}(r_0-r_{0\c}), \quad
u = \mu^{-\epsilon}A_{d}Z_u^{-1}Z_\phi^2u_0, \quad
\vec\phi = Z_\phi^{-1/2}\vec\phi_0, \label{eq:renormalization} 
\end{equation}
where in the minimal subtraction scheme to two-loop order 
\cite{zinn-justin89,amit84}
\begin{eqnarray}
Z_r(u,\epsilon) & = & 1 + 4(n+2)\frac{1}{\epsilon}u + (n+2)
\left[16(n+5)\frac{1}{\epsilon^2}-20\frac{1}{\epsilon}\right]u^2
+ O(u^3), \label{eq:Z_r} \\
Z_u(u,\epsilon) & = & 1 + 4(n+8)\frac{1}{\epsilon}u +
16 \left[ (n+8)^2\frac{1}{\epsilon^2} -(5n+22)\frac{1}{\epsilon}
\right]u^2 + O(u^3), \label{eq:Z_u}  \\
Z_\phi(u,\epsilon) & = & 1 - 4(n+2)\frac{1}{\epsilon}u^2
+ O(u^3). \label{eq:Z_phi}
\end{eqnarray}
In Eq.~(\ref{eq:renormalization}), $\mu^{-1}$ is a reference 
length and
\begin{equation}
A_d=\Gamma\left(1+ \epsilon/2 \right)
      \Gamma\left(1- \epsilon/2 \right) S_d \label{eq:ad}
\end{equation}
is a convenient geometrical factor where $\Gamma$ is Euler's gamma 
function. For applications to amplitude functions at $d<4$, this 
factor $A_d$ has considerable 
advantage \cite{dohm85,dohm85',schloms89,schloms90} over the more 
commonly used \cite{zinn-justin89,amit84} geometrical factor 
$S_d=[2^{d-1}\pi^{d/2}\Gamma(d/2)]^{-1}$.

\begin{figure}
\begin{picture}(400,50)
\put(160,25){\begin{picture}(100,40)
          \put(40,0){\circle{40}}
          \put(20,0){\circle*{5}}
          \put(60,0){\circle*{5}}
          \put(5,0){\line(30,0){70}}
          \end{picture} }
\end{picture}
\caption{Two-loop diagram of the two-point vertex function
$\Gamma_0^{(2)}$ above $T_\c$ whose dimensionally regularized
integral has a pole at $d=3$ as contained in 
Eqs.~(\ref{eq:Gamma^2(q)}), (\ref{eq:a2}) and 
(\ref{eq:pole_r0c}) of App.~A. This pole is also contained in 
similar diagrams of $\Gamma_0$ 
and $\Gamma_{0T}^{(2)}$ in Figs.~\protect\ref{fig:Gamma_0} and 
\protect\ref{fig:Gamma_T}.}
\label{fig:theta}
\end{figure}
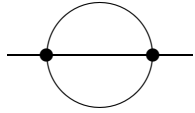

The pole at $d=3$ in $r_{0\c}$ can be traced to two-loop diagrams 
of the type shown in 
Fig.~\ref{fig:theta}. Having determined its coefficient 
(see Appendix A), we write
\begin{equation}
r_{0\c} = \frac{1}{\pi^2}(n+2)\frac{u_0^{2/\epsilon}}{\epsilon -1}
        +\tilde{S}(\epsilon,n)\, u_0^{2/\epsilon} \label{eq:r0C}
\end{equation}
where $\tilde{S}(\epsilon,n)$ is finite for $\epsilon=1$. 
As shown previously \cite{schloms89} an explicit knowledge of the 
function $\tilde{S}(\epsilon,n)$
is not necessary since $\tilde{S}(\epsilon,n)$ does not enter the 
final expressions of 
thermodynamic quantities as functions of the reduced temperature 
$t$. For convenience, therefore, we use instead of $r_0-r_{0\c}$ the
variable $r_0'=r_0-\delta r_0$
where $\delta r_0$ contains the $d=3$ pole of $r_{0\c}$ but not the 
poles of $\tilde{S}(\epsilon,n)$ at $d\not=3$. Thus we take
\begin{equation}
\label{eq:r0_pr}
r_0' = r_0 -\frac{1}{\pi^2}(n+2)\frac{u_0^{2/\epsilon}}{\epsilon -1}
           - C(n)\, u_0^{2/\epsilon} \label{eq:shift}
\end{equation}
with a conveniently chosen finite constant $C(n)$ whose
value is given in Eq.~(\ref{eq:C_eps}).
For $n=1$, this $r_0'$ coincides with the variable $r_0'$ used 
previously  \cite{bagnuls87,halfkann92}. The final results do 
not depend on the choice of $C(n)$.

When the perturbation series are expressed in terms of the
variables $r_0-r_{0\c}$ (or $r_0'$) the series at $d=3$
contain logarithms of $u_0$ arising from the nonanalytic $u_0$ 
dependence of $r_{0\c}$ (or $\delta r_0$). 
For higher-order calculations intended for resummation, 
such logarithms are inconvenient since the series are then 
not Borel-summable.
Perturbation series that are free of these logarithmic terms
(and which are also finite in $d=3$) can be obtained if they 
are instead
expressed in terms of the correlation length 
\cite{schloms89,schloms90} (or some other physical quantity).

Above $T_\c$, the correlation length $\xi_+$ is defined as usual by 
\begin{equation}
\xi_+^2 = \left. \bare{\chi}_+(0)\,
\partial\bare{\chi}_+(q)^{-1}/\partial q^2 
\right|_{q=0}
\label{eq:xi+2}
\end{equation}
where $\bare{\chi}_+(q)$ is the bare susceptibility at finite
wavenumber $q$.
Below $T_\c$, a common definition of the correlation length for 
$n=1$ and $n>1$ is complicated by the fact that, for $n>1$, the 
spatial decay of the order parameter correlations is not exponential 
\cite{privman91,fisher73,note1,pata73}.
Therefore, it is not straightforward to define an analogous quantity 
which plays the same role as $\xi_+$ above $T_\c$ in removing the 
logarithms in $u_0$ from the perturbation series obtained below 
$T_\c$.
Here, we follow Schloms and Dohm \cite{schloms90} who have introduced 
a pseudo-correlation length $\xi_-$ which performs precisely 
this task. We use Eq.~(\ref{eq:xi+2}) and the definition of $\xi_-$
given in Ref.~\cite{schloms90} to determine the two-loop expressions 
for $r_0' $ as 
a function of $\xi_+$ and $\xi_-$ in three dimensions. The results 
are (see Appendix~\ref{app:corrl})
\begin{eqnarray}
r_0'  = \xi_+^{-2} \Bigg\{ && 1 + (n+2)\left[
\frac{1}{\pi}u_0\xi_+ + \frac{1}{\pi^2} (u_0\xi_+)^2 \left( 
\frac{1}{27} +2 \ln(24u_0\xi_+) \right) \right] \nonumber\\
&& +\, O(u_0^3\xi_+^3) \Bigg\}\,, \label{eq:r0prime+}\\
-2r_0'  = \xi_-^{-2} \Bigg\{ && 1 + (n+2)\left[
\frac{1}{\pi}u_0\xi_- -\frac{1}{\pi^2} (u_0\xi_-)^2 \left( 
\frac{1385}{108} +4 \ln(24u_0\xi_-) \right) \right] \nonumber \\
&& +\, O(u_0^3\xi_-^3) \Bigg\} \label{eq:r0prime-}
\end{eqnarray}
for $T>T_\c$ and $T<T_\c$, respectively.
For $n=1$, Eqs.~(\ref{eq:r0prime+}) and (\ref{eq:r0prime-}) agree 
with 
Eqs.~(A11) and 
(3.8) of Halfkann and Dohm \cite{halfkann92}. These formulas
are not to be regarded as relations between the correlation length
and the reduced temperature $t$ but rather as an intermediate
step in the calculation of the bare quantities as functions of the 
correlation lengths $\xi_{\pm}$. The bare quantities can finally be 
obtained as functions of $t$ via the
connection between $\xi_\pm$ and $t$ as given in Appendix A.

For the purpose of deriving physical quantities as a
function of $\xi_\pm$ in three dimensions it is not nessecary to 
calculate the perturbation
series first as a function of $r_0'$ at $d=3$ and then to substitute 
$r_0'$ as a function of
$\xi_\pm$. A more direct way would be to start from (unevaluated) 
diagrammatic expressions
as a function of $r_0$ at $d\not=3$ and to substitute $r_0$ in the 
form of (unevaluated)
diagrammatic expressions as a function of $\xi_\pm$ at $d\not=3$. 
The resulting perturbation
series would consist of integral expressions that have both finite 
limits for $d\to3$ and are 
free of logarithms in $u_0$. The advantage of this procedure is that 
it avoids an explicit 
treatment of $d=3$ poles and it requires an evaluation only of a 
simplified form of finite 
integral expressions at $d=3$.
We have not chosen this more direct route of calculation in the 
present paper since at the
two-loop level its advantage is not yet substantial. In future 
calculations beyond two-loop
order, however, the simplification and reduction of computational 
labor implied by this
procedure may become crucial. For present purposes, the intermediate 
expressions given in terms
of $r_0'$ illustrate the appearence of the $\ln u_0$ terms and 
enable us to make contact with
the earlier work of Refs.~\cite{bagnuls87} and \cite{halfkann92}.

Finally, let us note that there is some flexibility in the 
definition of $\xi_-$ for general 
$n$. While it would be natural to define $\xi_-$ so as to coincide, 
for $n=1$, with the usual
correlation length of Ising-like systems below $T_\c$, a suitably 
modified definition of
$\xi_-$ for $n>1$ could well absorb the poles at $d<4$ and the 
logarithms in $u_0$ and yet
incorporate higher-order terms that are better adapted
to the region $T<T_\c$ than those of $\xi_-$ of our paper.
(The higher-order terms of our $\xi_-$ in Eq.~(\ref{eq:r0prime-}) 
are determined essentially from $\bare{\chi}_+(q)$ \em above \em 
$T_\c$, see Appendix A.)
A definition of $\xi_-$ for $n=1$ via the $q^2$ part of 
$\bare{\chi}_-(q)^{-1}$ could
formally be extended to $n>1$, for example, by including only 
those diagrammatic
contributions that cause an exponential spatial decay of the 
correlation function (i.e.,
by omitting those transverse parts of diagrams causing the 
algebraic decay).
This may lead to a simplified representation of the amplitude 
functions of other physical quantities below $T_\c$ at higher order.

%% file: free.tex
%
%
\section{Bare free energy}
\label{sec:free_2}
Because of our interest in computing the superfluid density for 
$^4$He ($n=2$), we consider a state in which the order parameter
$\langle\vec{\phi}_0({\bf x})\rangle$ has a uniform 
twist \cite{fisher73,rudnick77} along a fixed direction 
specified by a wave-vector ${\bf k}$.
For $n=2$, this twisted state is equivalent to the 
situation in $^{4}$He at constant superfluid velocity
${\bf v}_{\!s}=\hbar{\bf k}/m$ where the order parameter is a 
complex macroscopic wavefunction of plane wave structure 
$\langle\psi_0({\bf x})\rangle
=\eta_0(r_0,k)\exp\, (\i{\bf k\!\cdot\! x})$.
In three dimensions this case has been studied recently in 
one-loop order \cite{haussmann92}. 
Here, however, the situation is more delicate than in the earlier 
work on account of the (spurious) Goldstone divergences that 
plague the perturbation theory beyond one-loop order.
 
We begin by introducing, for $n\ge2$, an external field
\begin{equation}
\vec{h}_0({\bf x})=(h_0\cos{\bf k\!\cdot\! x},\, 
h_0\sin{\bf k\!\cdot\! x},\, 0,\ldots,0) \label{eq:h0_twist}
\end{equation}
which not only regulates the Goldstone divergencies but also 
generates the twist in the order
parameter; the amplitude $h_0$ is independent of \bf x\rm. 
At the same time, we introduce a local rotation according to
\begin{eqnarray}
\phi_{01}({\bf x}) &=& \phi_{01}'({\bf x})\,\cos{\bf k\!\cdot\! x}
           \,\, - \,\, \phi_{02}'({\bf x})\,\sin{\bf k\!\cdot\! x}
\, , \nonumber \\
\phi_{02}({\bf x}) &=& \phi_{01}'({\bf x})\,\sin{\bf k\!\cdot\! x}
           \,\, + \,\, \phi_{02}'({\bf x})\,\cos{\bf k\!\cdot\! x}
\, ,\nonumber \\
\phi_{0i}({\bf x}) &=& \phi_{0i}'({\bf x}) \, ,
\mbox{\hspace*{5cm}} (i\neq 1,2)\,. \label{eq:phi0_prime}
\end{eqnarray}
Substituting Eqs.~(\ref{eq:h0_twist}) and~(\ref{eq:phi0_prime}) 
into Eq.~(\ref{eq:LGW}), we obtain
{
\mathindent0pt
\begin{eqnarray}
\lefteqn{ {\cal H}\{\vec{\phi}_0({\bf x})\} = {\cal H}' 
\{\vec{\phi}_0'({\bf x),k} \}
= \int {\d}^dx \left[ \frac{1}{2}r_0\phi_0'^2 
+ \frac{1}{2}k^2(\phi_{01}'^2 +\phi_{02}'^2)
+ \frac{1}{2} \sum_{i=1}^n (\nabla\phi_{0i}')^2 
\right. }\nonumber\\
&& \phantom{{\cal H}\{\vec{\phi}_0({\bf x})\} = {\cal H}' \{ 
\vec{\phi}_0'({\bf x),k} \} = \int_V {\d}^dx \Bigg[}
+\left. {\bf k \cdot j} +u_0(\phi_0'^2)^2 -\vec{h}_0' \cdot 
\vec{\phi}_0' \right] \label{eq:LGW_prime}
\end{eqnarray}
}
where
${\bf j}({\bf x}) = \phi_{01}' ({\bf x})\nabla\phi_{02}' 
({\bf x}) - \phi_{02}' ({\bf x})\nabla\phi_{01}' ({\bf x}) $.
The advantage of working with the variables $\phi_{0i}' ({\bf x})$
is that $\vec{\phi}_0' ({\bf x})$ is coupled to the
spatially homogeneous field $\vec{h}_0' =(h_0,0,\ldots,0)'$ 
through $\vec{h}_0' \cdot \vec{\phi}_0'=h_0\phi_{01}'$.
We anticipate that the order parameter 
$\langle\vec{\phi}_0' \rangle=(M_0,0,\ldots,0)'$ in the rotated
system will also be homogeneous. In the original coordinate system, 
this corresponds to the twisted order parameter 
\cite{fisher73,rudnick77}
\begin{equation}
\langle\vec{\phi}_0({\bf x})\rangle=(M_0\cos{\bf k\!\cdot\! x},\,
M_0\sin{\bf k\!\cdot\! x},\, 0,\ldots,0) \label{eq:M0_of_x}
\end{equation}
where $M_0$ is independent of ${\bf x}$; 
the ${\bf k}$-dependent terms in Eq.~(\ref{eq:LGW_prime}) 
represent the 
additional energy associated with the applied twist. This is 
the same situation as studied previously 
\cite{haussmann92,fisher73,rudnick77} except
that now $M_0(r_0,k,h_0)$ depends not only on $r_0$ and $k$ but 
also on $h_0$.
Since the twist affects only the orientation of the order 
parameter in the
1-2 plane of its $n$-dimensional space, there are $(n-2)$ 
equivalent transverse components left. 
Their equivalence will be manifested through 
terms proportional to $(n-2)$ in the expression for the
free energy in Eq.~(\ref{eq:F_2loop}) and Fig.~\ref{fig:Gamma_0}.

\begin{figure}
\vspace*{-4cm}\hspace*{-26mm}\psfig{figure=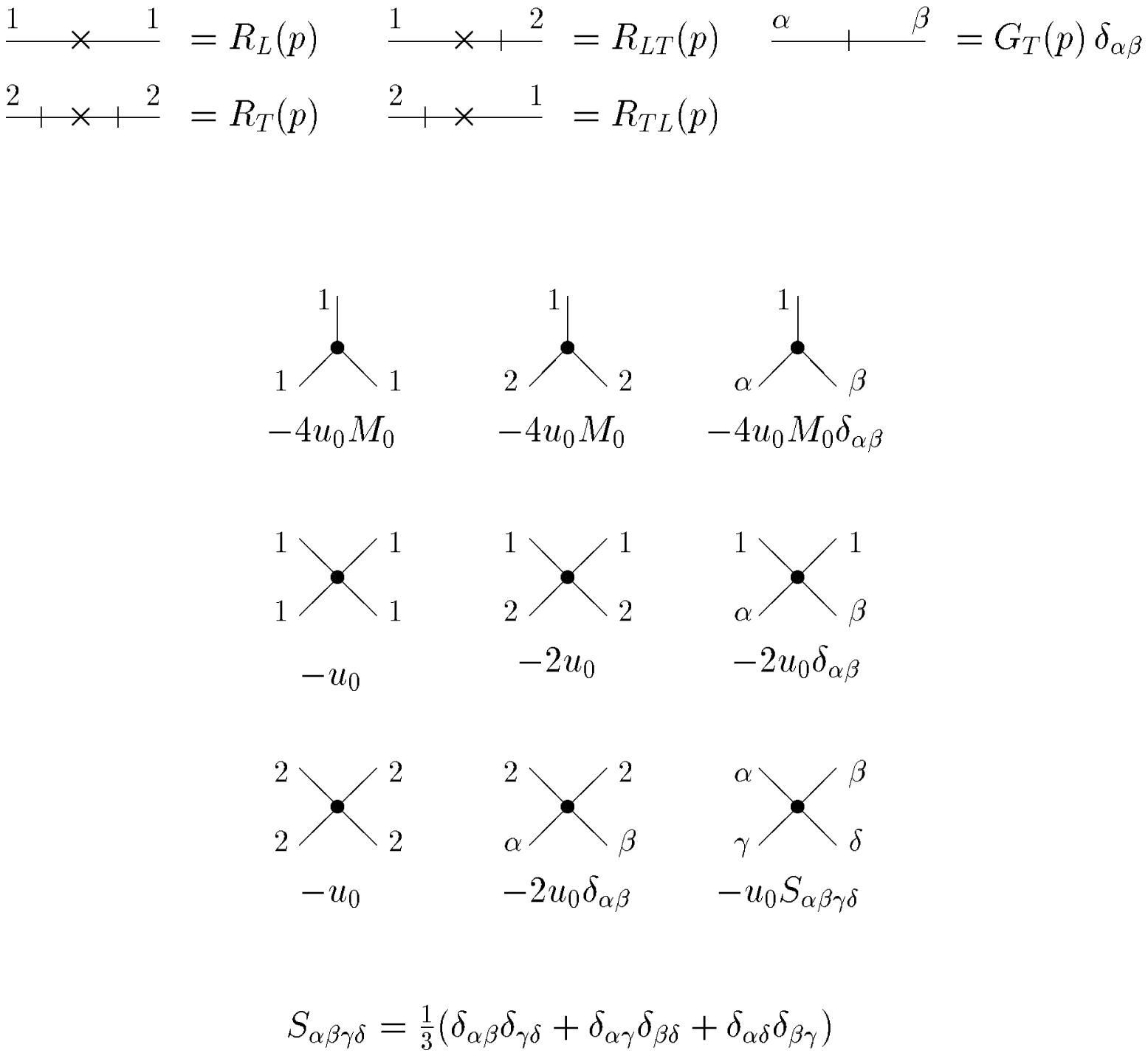,width=17cm}
\vspace{-87mm}
\caption{Propagators and vertices corresponding to 
Eqs.~(\ref{eq:LGW_prime-M0})--(\ref{eq:G_L}).
$\delta_{\alpha\beta}$ is the Kronecker delta and 
$\alpha\,,\beta\,,\gamma\,,\delta\,\,=\,3,4,\ldots, n$. For 
$R_L$, $R_T$, $R_{LT}$, $R_{TL}$ 
and $G_T$ see Eqs.~(\ref{eq:R_L}), (\ref{eq:R_LT}) and 
(\ref{eq:G_L}), respectively.}
\label{fig:vertices}
\end{figure}

\begin{figure}
\vspace*{-7mm}\hspace*{-28mm}\psfig{figure=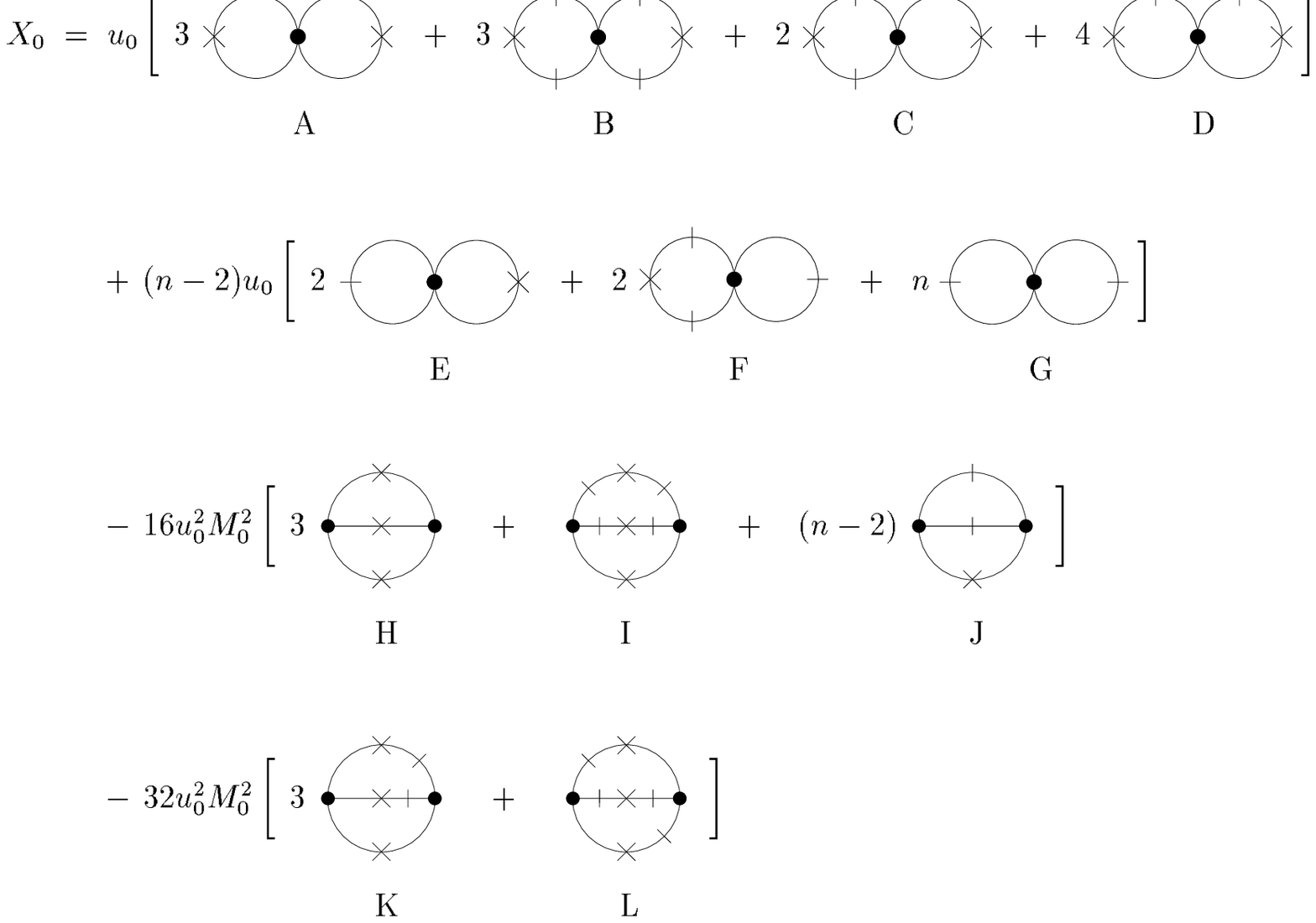,width=17cm}
\vspace*{-13.5cm}
\caption{Two-loop diagrams for the Helmholtz free energy 
$\Gamma_0$ in Eq.~(\ref{eq:F_2loop}) for a state with a twisted 
order-parameter, Eq.~(\ref{eq:M0_of_x}). The crosses $(\times)$
indicate $\bf k$ dependent propagators, compare 
Fig.~\ref{fig:vertices}. Correspondig integral
expressions for $X_{0i}$ are given in App.~C.
The $k^2$ parts of the diagrams C and I contain Goldstone 
divergencies [see Eqs.~(\ref{eq:g2c_3d_text}) and 
(\ref{eq:g2i_k_h_text})].} 
\label{fig:Gamma_0}
\end{figure}

We are now in a position to set up the perturbation theory. 
We make an expansion around
the exact average $\langle\vec{\phi}_0' \rangle$ according to
\begin{equation}
\vec{\phi}_0' ({\bf x}) = \langle\vec{\phi}_0' \rangle
+\delta\vec{\phi}_0' ({\bf x}) \label{eq:shift_M0}
\end{equation}
so that $\langle\delta\vec{\phi}_0' ({\bf x})\rangle=0$.
Substitution of 
Eq.~(\ref{eq:shift_M0}) into
Eq.~(\ref{eq:LGW_prime}) and use of the Fourier representation
\begin{equation}
\delta\vec{\phi}_0' ({\bf x}) = \int_{{\bf p}}
\delta\vec{\varphi}_0' ({\bf p}) e^{\i{\bf p \!\cdot\! x}}
\label{eq:fourier}
\end{equation}
where $\int_{\mathbf p}\equiv(2\pi)^{-d}\int{\d}^dp$, leads to
\begin{eqnarray}
{\cal H}' \{\vec{\phi}_0' ({\bf x),k}\}
&=& {\cal H}' \{ \langle\vec{\phi}_0' \rangle,{\bf k} \} + 
\int{\d}^dx \left.
\frac{\delta{\cal H}' \{\vec{\phi}_0' ,{\bf k}\}}{\delta 
\vec{\phi}_0' ({\bf x})}
\right|_{\vec{\phi}_0' =\langle\vec{\phi}_0' \rangle}
\mbox{\hspace*{-3mm}}\cdot\delta\vec{\phi}_0' ({\bf x}) \nonumber\\
&& +\, \frac{1}{2}\int_{\bf p} \delta\vec{\varphi}_0' (-{\bf p})
\tilde{K}({\bf p,k}) \delta\vec{\varphi}_0' ({\bf p}) \nonumber\\
&& +\, \int {\d}^dx \left[ 4u_0(\langle\vec{\phi}_0' 
\rangle \cdot\delta\vec{\phi}_0' )
\delta\phi_0^{\prime 2} +u_0(\delta\phi_0^{\prime 2})^2
-\vec{h}_0' \cdot\delta\vec{\phi}_0' \right]     
\label{eq:LGW_prime-M0}
\end{eqnarray}
where $\tilde{K}$ is the $n\times n$ matrix
\begin{equation}
\tilde{K}({\bf p,k})= \left( \begin{array}{cc}
    K({\bf p,k}) & 0 \\[3pt]
    0 & \mbox{\hspace*{3mm}}(\bar{r}_{0T}+p^2) I_{n-2}
    \end{array} \right). \label{eq:G_p_k}
\end{equation}
$I_{n-2}$ is the $(n-2)\times (n-2)$ unit matrix and
\begin{eqnarray}
K({\bf p,k}) =
\left( \begin{array}{cc}
\bar{r}_{0L}+k^2+p^2 \mbox{\hspace*{3mm}} & 2i\,{\bf k\!\cdot\! p} 
\\
-2i\,{\bf k\!\cdot\! p} \mbox{\hspace*{3mm}}
& \bar{r}_{0T}+k^2+p^2 \end{array} \right)\,, 
\label{eq:matrix_K} \\[4mm]
\bar{r}_{0L} = r_0 + 12u_0M_0^2\,, \quad 
\bar{r}_{0T} = r_0 +  4u_0M_0^2    \label{eq:r0L} 
\end{eqnarray}
(see also Ref.~\cite{haussmann92}).
For $k=0$, the diagonal elements of $K({\bf p,0})^{-1}$ yield 
the standard longitudinal and transverse propagators
\begin{equation}
G_L(p) = (\bar{r}_{0L}+p^2)^{-1}\,, \quad
G_T(p) = (\bar{r}_{0T}+p^2)^{-1}\,. \label{eq:G_L}
\end{equation}
The diagrammatic rules corresponding to
Eqs.~(\ref{eq:LGW_prime-M0})--(\ref{eq:G_L}) are indicated in 
Fig.~\ref{fig:vertices}. 
The difference between the present perturbation theory and 
that for ${k=0}$ case lies in the ${\bf k}$-dependence of the 
propagators and in the greater number of vertices due to
the three (rather than two) types of components for the order 
parameter. 
The resulting two-loop expression for the (bulk) Helmholtz 
free energy 
per unit volume reads (up to an unimportant additive constant)
\begin{eqnarray}
\lefteqn{ \Gamma_0(r_0,u_0,M_0,k) =
\frac{1}{2}(r_0+k^2)M_0^2+u_0M_0^{4} 
+\frac{1}{2}\int_{{\bf p}} \ln \left[\det K({\bf p,k})\right] } 
\hspace{1cm} \nonumber\\
&& +\, \frac{1}{2}(n-2)\int_{{\bf p}}\ln (\bar{r}_{0T}+p^2) 
+ X_0(r_0,u_0,M_0,k) +O(u_0^2) \label{eq:F_2loop}
\end{eqnarray}
where $X_0$ is considered to be of $O(u_0)$ and denotes the negative
sum of the 1PI two-loop vacuum diagrams shown in 
Fig.~\ref{fig:Gamma_0}.
We label these diagrams A, B, \ldots , L and refer to the 
associated integral 
expressions as $X_{0A}$, $X_{0B}$, \ldots , $X_{0L}$ [see App.~C].
We have omitted diagrams with tadpole insertions since their sum, 
being equal to $\langle\delta\vec{\phi}_0' ({\bf x})\rangle$, 
vanishes according to Eq.~(\ref{eq:shift_M0}). 
Eq.~(\ref{eq:F_2loop}) provides the basis for deriving the 
helicity modulus in Sect.~\ref{sec:rho_s}. For $k=0$,
Eq.~(\ref{eq:F_2loop}) reduces to the two-loop expression given 
by Chang and Houghton \cite{chang80b} and by Shpot \cite{shpot90}.

Like the integral associated with Fig.~1, the integrals 
$X_{0H}$, $X_{0I}$ and $X_{0J}$ are ultraviolet divergent in 
three dimensions and thus have poles at $\epsilon=1$.
We denote these integrals by $\bare{X}_H$, $\bare{X}_I$ and 
$\bare{X}_J$ when they are expressed
as functions of $r_0'$, Eq.~(\ref{eq:shift}). 
In order to collect all $d=3$ pole terms of 
$\Gamma_0(r_0,u_0,M_0,0)$ we consider
$\bare{X}_H$, $\bare{X}_I$ and $\bare{X}_J$ at $k=0$ and define 
\begin{equation}
u_0^2M_0^2\, f(r_0' ,u_0,M_0)=\lim_{\epsilon\rightarrow 1}
\left[\frac{1}{2}M_0^2\delta r_0 + \bare{X}_H
 + \bare{X}_I + \bare{X}_J \right]_{k=0} \label{eq:f}
\end{equation}
where
\begin{eqnarray}
\lefteqn{ \left[ \bare{X}_H+\bare{X}_I+\bare{X}_J \right]_{k=0} =
-16u_0^2M_0^2 \frac{1}{\epsilon-1} r_{0L}^{1-\epsilon}
            \frac{\Gamma(\epsilon)}{(4\pi)^{d}} I(\epsilon)\,,} 
\hspace{2cm} \label{eq:G2H+I+J_DR} \\
I(\epsilon) &=&
\int_0^{1}\mbox{d}x \int_0^{1}\mbox{d}y\: y^{\epsilon/2-1}\:
\frac{3 +(n-1) [1 -y\,(1-r_{0T}/r_{0L})]^{1-\epsilon}}{[\,
1-y+y(x-x^2)\,]^{2-\epsilon/2}}\,, \label{eq:J1} \\
r_{0L} &=& r_0' +12u_0M_0^2\,, \quad 
r_{0T} = r_0' +4u_0M_0^2\,,  \label{eq:r0Lprime}
\end{eqnarray}
with $I(1)=2\pi(n+2)$. It is understood that
$r_0$ is replaced by $r_0' $ everywhere else in
Eq.~(\ref{eq:F_2loop}), the difference affecting only terms of 
higher order. Expanding $\bare{X}_H$, $\bare{X}_I$ and 
$\bare{X}_J$ around $\epsilon=1$, one finds that the pole term in
Eq.~(\ref{eq:G2H+I+J_DR}) is cancelled by the pole of
$\frac{1}{2}M_0^2\delta r_0$.
The resulting quantity $f$ is given by
\begin{eqnarray}
f(r_0',u_0,M_0) &=& 
\frac{3}{\pi^2} \ln \frac{r_{0L}^{1/2}}{24u_0}
+\frac{n-1}{\pi^2} \ln \frac{r_{0L}^{1/2}+2r_{0T}^{1/2}}{72u_0}\,. 
\label{eq:f_r0}
\end{eqnarray}
The contributions $\bare{X}_D$,
$\bare{X}_K$ and $\bare{X}_L$ vanish at $k=0$ and the 
remaining two-loop contributions reduce to products of standard 
one-loop integrals. 
Thus, we obtain for the bare Helmholtz free energy per unit 
volume in two loop order at $k=0$ and $d=3$
\begin{eqnarray}
\bare{\Gamma}(r_0' ,u_0,M_0) &=& 
\frac{1}{2}r_0' M_0^2+u_0M_0^{4}
-\frac{1}{12\pi}r_{0L}^{3/2} -\frac{1}{12\pi}(n-1)r_{0T}^{3/2}
\nonumber \\ & + & \mbox{} \frac{3}{16\pi^2}u_0r_{0L}
 +\frac{1}{8\pi^2}(n-1)u_0r_{0L}^{1/2}r_{0T}^{1/2} 
 +\frac{1}{16\pi^2}(n^2-1)u_0r_{0T}
\nonumber \\ & + &  \mbox{} u_0^2M_0^2\, f(r_0' ,u_0,M_0)
+ O(u_0^2,u_0^2\ln u_0) \, . \label{eq:free_k=0}
\end{eqnarray}
For the case $n=1$, this expression reduces to the two-loop part of 
Eq.~(3.18) of Bagnuls et al.~\cite{bagnuls87}
and of Eq.~(3.3) of Halfkann and Dohm \cite{halfkann92} 
and agrees with their coefficients $F_{blk}$ for $b=1$ and $b=2$.
The latter agree also with those in Eq.~(A1.1) of Guida and 
Zinn-Justin \cite{guida96} for $n=1$.
Eq.~(\ref{eq:free_k=0}) contains logarithms of the coupling via 
$u_0^2M_0^2f(r_0',u_0,M_0)\sim O(u_0\ln u_0)$ as expected.
Unlike the case $n=1$, however, the two terms proportional to $n-1$ 
in Eq.~(\ref{eq:free_k=0}) depend nonanalytically on $r_{0T}$ 
through $r_{0T}^{3/2}$ and $r_{0T}^{1/2}$.
These nonanalyticities lead to perturbative terms in the 
derivatives of the free energy (with respect to $M_0$) 
that diverge when $r_{0T}\rightarrow 0$. The origin of
these divergences are the well-known Goldstone modes 
\cite{goldstone61} that result from the fact that, 
for a bulk homogeneous system
with a rotationally symmetric order parameter, there is no restoring
force to a global rotation of the order parameter at $h_0=0$ 
\cite{pata73}.
For the rotationally invariant quantities that we consider 
here (square of the order parameter, specific heat, 
stiffness constant), the divergences of
individual perturbative contributions should be spurious 
and should cancel among themselves leaving a finite result 
in the limit $h_0\rightarrow 0$,
as required on general grounds \cite{lawrie81,elitzur83}.
Within our two-loop calculation, we shall show that this is 
indeed the case. 
This is in contrast to the transverse and longitudinal
susceptibilities which possess true physical singularities in this 
limit \cite{pata73,lawrie81,brezin73,nelson76,schaefer78}.

%% file: M0.tex
%
%
\section{Order parameter}
\label{sec:M0}
\subsection{Bare theory}
The bare perturbative expression for the order parameter 
$M_0$ at $k=0$ can be obtained either by evaluating the diagrams 
\cite{shpot90,bervillier76}
of the one-point vertex function $\Gamma_0^{(1)}$
(see Appendix~\ref{app:eq_state})
or, as we do here, by differentiation of the bare free 
energy $\bare{\Gamma}$, Eq.~(\ref{eq:free_k=0}), according to
\begin{equation}
h_0(r_0',u_0,M_0)=
\frac{\partial}{\partial M_0}\bare{\Gamma}(r_0',u_0,M_0)\,.
\label{eq:state}
\end{equation}
In three dimensions, we obtain
\begin{eqnarray}
\lefteqn{\frac{1}{M_0}h_0(r_0',u_0,M_0) =
r_0'+4u_0M_0^2 -\frac{3}{\pi}u_0r_{0L}^{1/2}  - 
\frac{1}{\pi}(n-1)u_0r_{0T}^{1/2} }\nonumber\\
&+& \frac{3}{2\pi^2}(n-1)u_0^2 \left[ 
\left( \frac{r_{0T}}{r_{0L}} \right)^{-1/2} \!+ \left( 
\frac{r_{0T}}{r_{0L}} \right)^{1/2}
\!- \frac{r_{0T}}{r_{0L}+2(r_{0T}r_{0L})^{1/2}} \right] \nonumber\\
&+& 2u_0^2 f(r_0',u_0,M_0)
+ \frac{1}{2\pi^2}u_0^2 \left[ n^2-n+18 - 9\frac{r_{0T}}{r_{0L}} 
\right] +O(u_0^3,u_0^3\ln u_0)\,.  \label{eq:h0tilde}
\end{eqnarray}
From the two-loop term $\sim r_{0T}^{-1/2}$, it appears
that this perturbative expression, in its present form, is
problematic for small $r_{0T}$. In particular, 
Eq.~(\ref{eq:h0tilde}) cannot be used directly at $h_0=0$ 
to obtain $M_0$.
To circumvent this problem we invert Eq.~(\ref{eq:h0tilde}) 
iteratively at $h_0\neq 0$. This leads to the bare perturbative 
form of the (implicit)
equation of state (for $M_0^2$ as a function of $r_0'$ and $h_0$) 
in three dimensions
\begin{eqnarray}
M_0^2 &=& \frac{1}{4u_0} (-r_0'+\bare{\chi}_{{}_T}^{-1})
+\frac{3}{4\pi} (-2r_0'+3\bare{\chi}_{{}_T}^{-1})^{1/2} 
+\frac{1}{4\pi}(n-1) \bare{\chi}_{{}_T}^{-1/2} \nonumber\\
&& +\, \frac{1}{8\pi^2}(n-1)u_0 \left[ 6w^{1/2} +3w(1+2w^{1/2})^{-1} 
-4\ln \frac{1+2w^{1/2}}{3} \right] \nonumber\\
&& +\, \frac{1}{8\pi^2} u_0 (10-n+9w) - \frac{1}{2\pi^2}(n+2)u_0 
\ln \frac{(-2r_0'+3\bare{\chi}_{{}_T}^{-1})^{1/2}}{24u_0} 
\nonumber\\
&& +\, O(u_0^2,u_0^2\ln u_0) \,, \label{eq:M02_of_h} \\[2mm]
w &=& \bare{\chi}_{{}_T}^{-1} (-2r_0'+3\bare{\chi}_{{}_T}^{-1})^{-1}
\label{eq:w,lambda}
\end{eqnarray}
where $h_0$ enters via the transverse susceptibility
\begin{equation}
\bare{\chi}_{{}_T} = M_0 / h_0 \label{eq:chiT}
\end{equation}
and $w=O(h_0)$ for $h_0\to0$ and $r_0'<0$.

Eq.~(\ref{eq:h0tilde}) also yields the bare longitudinal 
susceptibility 
$\bare{\chi}_{{}_L}(r_0',u_0,M_0)$ in two-loop order since
\begin{equation}
\bare{\chi}_{{}_L}^{-1}(r_0',u_0,M_0) = 
\frac{\partial}{\partial M_0} h_0(r_0',u_0,M_0)\,. \label{eq:XLdef}
\end{equation}
Combining this with Eq.~(\ref{eq:M02_of_h}), we find the 
perturbative form of $\bare{\chi}_{{}_L}^{-1}$ in three dimensions
{
\mathindent0pt
\begin{eqnarray}
\bare{\chi}_{{}_L}^{-1} &=& -2r_0'+3\bare{\chi}_{{}_T}^{-1}
- \frac{1}{2\pi} u_0 (-2r_0'+3\bare{\chi}_{{}_T}^{-1})^{1/2} 
\left[ (n-1)w^{-1/2} (1-5w) -3-9w \right] \nonumber\\
&+& \frac{1}{4\pi^2}u_0^2 \left[ (n-1)^2w^{-1}+8(n-1)w^{-1/2}
+ (36 -4n -5n^2) +W(w) \right] \nonumber \\ 
&-& \frac{4}{\pi^2} u_0^2 \left[ (n+2) \ln 
\frac{(-2r_0'+3\bare{\chi}_{{}_T}^{-1})^{1/2}}{24u_0} - 
(n-1) \ln3 \right] +O(u_0^3,u_0^3\ln u_0) \label{eq:XL_of_h}
\end{eqnarray}
}
where the function
\begin{eqnarray}
\label{eq:W}
W(w) &=& \frac{3w^{1/2}}{(1+2w^{1/2})^2}\,
\Big[\, 4(n-1) +(8n-11)w^{1/2} -6(n+1)w -30n\, w^{3/2} \nonumber\\
&& -(30n+42)w^2 -72w^{5/2} \,\Big] -16(n-1) \ln ( 1+2w^{1/2} )
\end{eqnarray}
vanishes for $w=0$.
Note that the one- and two-loop terms in Eq.~(\ref{eq:XL_of_h}) 
diverge as $w^{-1/2}$ and 
$w^{-1}$ for $h_0\to0$ below $T_\c$. This apparent breakdown 
of perturbation theory signals
the Goldstone singularity of the longitudinal
susceptibility \cite{brezin73,nelson76,schaefer78}
in this limit. An appropriate description of this singularity 
near $T_\c$ is nontrivial \cite{lawrie81} and will be 
studied elsewhere.

The behavior of $\bare{\chi}_{{}_L}$ should be contrasted 
with that for the order parameter $M_0$. 
In deriving Eq.~(\ref{eq:M02_of_h}) from Eq.~(\ref{eq:h0tilde}) 
or, equivalently, from the vertex function $\Gamma_0^{(1)}$ 
(see App.~B) we encounter terms below $T_\c$ that diverge as 
$\bare{\chi}_{{}_T}^{1/2}\sim h_0^{-1/2}$ 
for $h_0\to0$ but which cancel among themselves.
A similar cancellation of $\ln \bare{\chi}_{{}_T}$ terms near 
four dimensions was noted
previously by Br\'ezin, Wallace and Wilson \cite{brezin(2)73}.
In three dimensions, we obtain from Eq.~(\ref{eq:M02_of_h}) 
the finite result
\begin{eqnarray}
M_0^2 &=& \frac{1}{8u_0}(-2r_0')+\frac{3}{4\pi}(-2r_0')^{1/2} 
-\frac{1}{2\pi^2}(n+2) u_0\ln \frac{(-2r_0')^{1/2}}{24u_0} 
\nonumber\\
&& +\, \frac{1}{8\pi^2} u_0 \left[ 10 -n +4(n-1)\ln 3 \right]
+O(u_0^2,u_0^2\ln u_0) \label{eq:M0_r0prime}
\end{eqnarray}
for $r_0'<0$ and $h_0\to0$.
The final step is to rewrite of Eq.~(\ref{eq:M0_r0prime}) 
in terms of the correlation length $\xi_{-}$ in order to remove 
the logarithms in $u_0$. Using Eq.~(\ref{eq:r0prime-}) 
for $r_0'$ as a function of $\xi_{-}$,
we obtain the following two-loop expression for the square of 
the bare order parameter at $h_0=0$
\begin{eqnarray}
M_0^2(\xi_-,u_0,3) = \xi_-^{-1}\, \Bigg\{ && \frac{1}{8u_0\xi_-} 
+ \frac{1}{8\pi}(n+8) -\left[ \frac{1}{864\pi^2}(1169n+1042) 
\right. \nonumber\\
&& - \left. \frac{1}{2\pi^2} (n-1)\ln 3 \right] u_0\xi_- 
+O(u_0^2\xi_-^2) \,\Bigg\} \label{eq:M0_xi-}
\end{eqnarray}
in three dimensions. For $n=1$ this agrees with Eq.~(3.13) and 
Table 2 of Halfkann and Dohm \cite{halfkann92}.
\subsection{Amplitude function of $M_0^2$ in two-loop order}
\label{ssec:M0_ampl}
The square of the renormalized
order parameter $M^2$ is obtained from $M_0^2(\xi_-,u_0,d)$ 
according to \cite{schloms90}
\begin{eqnarray}
\label{eq:M2}
M^2(\xi_-,u,\mu,d) &=& Z_{\phi}^{-1}M_0^2(\xi_-,
\mu^\epsilon A_d^{-1} Z_u Z_\phi^{-2}u,d) \\
&=& \xi_-^{2-d} f_\phi(\mu\xi_-,u,d)
\label{eq:def_fphi}
\end{eqnarray}
where $f_\phi$ is the amplitude function and where the 
renormalized parameters are defined in 
Eqs.~(\ref{eq:renormalization})--(\ref{eq:ad}). 
Solution of the RG equation for $M^2$ implies the following
decomposition of the bare order parameter in three dimensions
\begin{eqnarray}
\label{eq:flow_fphi}
M_0^2(\xi_-,u_0,3) &=& Z_\phi(u,1)\, \xi_-^{-1}\, 
f_\phi(1,u(l_-),3) \exp \left[-\int_u^{u(l_-)} 
\frac{\zeta_\phi(u')}{\beta_u(u',1)} {\mathrm d}u' \right]\,, \\
\zeta_\phi(u) &=& \left. \mu \partial_\mu \ln Z_\phi^{-1} 
\right|_0\,, \label{eq:zeta_r}\\
\beta_u(u,\epsilon) &=& -\epsilon u + \tilde{\beta}(u) = u 
\left[ -\epsilon + \left. \mu \partial_\mu \ln\, 
(Z_u^{-1}Z_\phi^2) \right|_0 \,\right]
\label{eq:beta_u}
\end{eqnarray}
where $\zeta_\phi$ and $\beta_u$ are
the well known RG functions \cite{schloms89}.
The flow parameter $l_-$ is chosen according to $l_-\mu\xi_-=1$ 
and the effective coupling
$u(l)$ is determined by Eq.~(\ref{eq:RG_u_of_l}).
We recall that the exponential part in Eq.~(\ref{eq:flow_fphi}) 
is determined, within the
minimal subtraction scheme, entirely by the $d=4$ poles carried 
by $Z_\phi$ and $Z_u$ and that the
renormalization of $u_0$ likewise involves only the pure poles 
contained within the product
$Z_uZ_\phi^{-2}$. Since these poles are the same above and below 
$T_\c$, we may use the
$Z$-factors as given in Eqs.~(\ref{eq:Z_r})--(\ref{eq:Z_phi}), 
without further calculation,
to determine the amplitude function. In three dimensions, 
$f_\phi$ is given by
\begin{equation}
\label{eq:f_phi(1,u,3)}
f_\phi(1,u,3) = Z_\phi^{-1}\, \xi_-\, M_0^2(\xi_-,4\pi \xi_-^{-1} 
Z_uZ_\phi^{-2}u,3)
\end{equation}
where we have used
\begin{equation}
A_3 = (4\pi)^{-1} \label{eq:A_3=1/4pi}
\end{equation}
and where $Z_\phi(u,1)$ and $Z_u(u,1)Z_\phi(u,1)^{-2}$ are given 
by Eqs.~(\ref{eq:Z_u}) and
(\ref{eq:Z_phi}) evaluated at $\epsilon=1$. Using 
Eq.~(\ref{eq:M0_xi-}) we obtain from Eq.~(\ref{eq:f_phi(1,u,3)})
\begin{equation}
f_{\phi}(1,u,3)=\frac{1}{32\pi u}+
\left[ \frac{1}{27\pi}(160-82n) 
+ \frac{2}{\pi}(n-1)\ln3 \right]u + O(u^2)
\label{eq:f_phi2}
\end{equation}
for the amplitude function in two-loop order. 
For $n=1$, this agrees with Eq. (3.21) and Table 3 of Halfkann 
and Dohm \cite{halfkann92}. Note that the vanishing of the 
one-loop term in Eq.~(\ref{eq:f_phi2}) is due to the choice 
of the geometric factor $A_{d}$
in the definition of the renormalized coupling \cite{schloms90}
in Eq.~(\ref{eq:renormalization}).

%% file: C.tex
%
%
\section{Specific heat}
\label{sec:C}
\subsection{Bare theory}
 Although the two-loop results for the amplitude functions 
of the specific heat (at $h_0=0$) have been given previously 
\cite{dohm85,schloms90},
we present here, for completeness, their derivation within the 
present approach. 
The bare specific heat $\bare{C}^\pm$ above (+) and below ($-$) 
$T_\c$ is given by
\begin{equation}
\bare{C}^\pm = C_{B} - a_0^2\,
\bare{\Gamma}_\pm^{(2,0)} \label{eq:C_G}
\end{equation}
where $C_{B}$ is a noncritical background value and
$a_0$ is a constant defined by $(r_0-r_{0\c})=a_0t$. 
The vertex functions $\bare{\Gamma}_{\pm}^{(2,0)}$ can be obtained
either from the 
sum of 1PI vacuum diagrams with two $\phi_0^2$ insertions or 
by differentiation of the free energy, Eq.~(\ref{eq:free_k=0}), 
according to
\begin{equation}
\bare{\Gamma}_\pm^{(2,0)}
= \frac{\d^2}{({\mathrm d}r_0')^2}\,
\bare{\Gamma}(r_0',u_0,M_0(r_0',u_0))\,. \label{eq:G_h=0}
\end{equation}
Using Eqs.~(\ref{eq:free_k=0}) and (\ref{eq:M0_r0prime}) we obtain
\begin{eqnarray}
\bare{\Gamma}_+^{(2,0)} & = &  
-\frac{n}{16\pi}r_0^{\prime \,-1/2}
+ O(u_0^2,u_0^2\ln u_0)\, , \label{eq:G20+} \\
\bare{\Gamma}_-^{(2,0)} & = & -\frac{1}{8u_0}
-\frac{1}{4\pi}(-2r_0')^{-1/2} +\frac{1}{4\pi^2}(n+2)u_0(-2r_0')^{-1}
+O(u_0^2,u_0^2\ln u_0) \nonumber\\
\label{eq:G20-}
\end{eqnarray}
in three dimensions for $r_0'>0$ and $r_0'<0$, respectively. 
As expected, Eqs.~(\ref{eq:G20+}) and~(\ref{eq:G20-}) contain 
no logarithms
in $u_0$ since, at this order, there are no diagrams of 
$\bare{\Gamma}_\pm^{(2,0)}$ with $d=3$ poles. Nevertheless, in
anticipation of future Borel summations, we rewrite the above 
expressions for $\bare{\Gamma}^{(2,0)}_\pm$ in terms of the 
correlation lengths $\xi_{\pm}$ in 
Eqs.~(\ref{eq:r0prime+}) and (\ref{eq:r0prime-}). Thus, we obtain
\begin{eqnarray}
\bare{\Gamma}_+^{(2,0)}(\xi_+,u_0,3) &=& \xi_+ \left[
-\frac{n}{16\pi} +\frac{n}{32\pi^2}(n+2)u_0\xi_+ + O(u_0^2\xi_+^2) 
\right]\,, \label{eq:C_xi+}  \\
\bare{\Gamma}_-^{(2,0)}(\xi_-,u_0,3) &=& \xi_- \left[ 
-\frac{1}{8u_0\xi_-} -\frac{1}{4\pi} +\frac{3}{8\pi^2}(n+2)u_0\xi_- 
+ O(u_0^2\xi_-^2) \right]
\label{eq:C_xi-}
\end{eqnarray}
in three dimensions. 
In an earlier calculation \cite{dohm85} of the specific heat in 
three dimensions below $T_\c$ (performed on the basis of the 
static distribution function of model C \cite{halperin74}) 
Goldstone singularities of individual diagrams were found to 
cancel among themselves, in accord with Eq.~(\ref{eq:G20-}).
\subsection{Amplitude functions of $\bare{C}^\pm$ in two-loop order}
The renormalized vertex functions $\Gamma_{\pm}^{(2,0)}$ are 
obtained from $\bare{\Gamma}_\pm^{(2,0)}(\xi_\pm,u_0,d)$ 
according to \cite{schloms90}
\begin{equation}
\Gamma_{\pm}^{(2,0)}(\xi_{\pm},u,\mu,d) = Z_r^2\,
\bare{\Gamma}_\pm^{(2,0)}(\xi_{\pm},
       \mu^{\epsilon}Z_uZ_\phi^{-2}A_d^{-1}u,d)\,
 -\, \frac{1}{4}\mu^{-\epsilon}A_dA(u,\epsilon) \label{eq:ren_G20}
\end{equation}
where, in two-loop order, the RG function
\begin{equation}
A(u,\epsilon)=-2n\frac{1}{\epsilon}-8n(n+2)\frac{1}{\epsilon^2}u
+O(u^2) \label{eq:A}
\end{equation}
absorbs the additive poles in both $\bare{\Gamma}_+^{(2,0)}$ and
$\bare{\Gamma}_-^{(2,0)}$ (see Eqs.~(3.20) and (5.9) of 
Ref.~\cite{dohm85} and Eq.~(2.18) of Ref.~\cite{krause90}). 
Dimensionless amplitude functions $F_{\pm}$ can then be defined 
according to
\begin{equation}
\Gamma_{\pm}^{(2,0)}(\xi_{\pm},u,\mu,d)=
-\frac{1}{4}\mu^{-\epsilon}A_dF_{\pm}(\mu\xi_{\pm},u,d) 
\label{eq:Fpm}
\end{equation}
so that the quantities of interest $F_\pm(1,u,3)$ are given by
\begin{equation}
\label{eq:F_+-}
F_\pm(1,u,3) = -16\pi Z_r^2\, \xi_\pm^{-1}\, 
\bare{\Gamma}_\pm^{(2,0)} (\xi_\pm,4\pi\xi_\pm^{-1} 
Z_u Z_\phi^{-2}u,3)\, +\, A(u,1)\,.
\end{equation}
Use of Eqs.~(\ref{eq:C_xi+}) and~(\ref{eq:C_xi-}) together with 
the $Z$-factors $Z_r(u,1)$, $Z_u(u,1)$,
$Z_\phi(u,1)$ and the additive function $A(u,1)$ of Eq.~(\ref{eq:A}) 
leads to the two-loop formulas for the amplitude functions
in three dimensions \cite{dohm85,schloms90}
\begin{eqnarray}
F_+(1,u,3) & = &  
-n -2n(n+2)u +O(u^2)\, , \label{eq:F+2loop} \\
F_-(1,u,3) & = & \frac{1}{2u}-4+8(10-n)u+O(u^2)\,.
 \label{eq:F-2loop}
\end{eqnarray}
Note that the first two terms of $F_\pm(1,u,3)$ remain valid 
for general $d\leq4$ due to the choice of the geometric factor 
$A_d$ in the definition of the renormalized coupling
\cite{dohm85}.

%% file: rho_s.tex
%
%
\section{Helicity modulus and superfluid density 
(stiffness constant)}
\label{sec:rho_s}
\subsection{Bare theory}
The helicity modulus $\Upsilon$ measures the change of the 
free energy resulting from a small
twist imposed on the order parameter \cite{fisher73,rudnick77}. 
This can be obtained from the change in the bare Gibbs free 
energy per unit volume $F_0(r_0,u_0,h_0,k)$ [see 
Eq.~(\ref{eq:gibbs})] caused by the $\mathbf k$ dependence of
the field described by Eq.~(\ref{eq:h0_twist}). For small $k$, 
this change is \cite{fisher73,rudnick77}
\begin{equation}
\Delta F_0 \equiv F_0(r_0,u_0,h_0,k)-
F_0(r_0,u_0,h_0,0)= \half \Upsilon k^2 + O(k^4)\,. 
\label{eq:DF}
\end{equation}
We are interested primarily in the physical case at $h_0=0$
for superfluid $^{4}$He $(n=2)$. In the presence of a uniform 
superfluid velocity ${\bf v}_{s}=\hbar{\bf k}/m$ 
(relative to the motion of the normal fluid), 
phenomenological considerations within the two-fluid model identify
the additional energy density as the kinetic energy density
$\frac{1}{2}\rho_{s}v_{s}^2  = k_{B}T\Delta F_0$.
This implies that, for small $k$ and $h_0=0$, the superfluid 
density $\rho_{s}$ should be related to $\Upsilon$ by 
\cite{fisher73}
\begin{equation}
\rho_{s}= k_{B}T\left( m/\hbar \right)^2\,\Upsilon\,. 
\label{eq:rho_s}
\end{equation}
For O($n$) symmetric magnetic systems $\rho_s$ is the stiffness 
constant \cite{privman91,hohenberg76}.

It turns out that at two-loop order
our perturbative treatment is plagued by spurious Goldstone 
singularities for $h_0=0$, $T<T_\c$. 
To circumvent this defect of perturbation theory we shall first 
work at $h_0\neq 0$ and define the helicity modulus as
\begin{equation}
\Upsilon= \left. 2\frac{\partial}{\partial k^2} 
F_0(r_0,u_0,h_0,k)\right|_{k=0}. \label{eq:UPs}
\end{equation}
We shall see that the Goldstone singularities are cancelled 
in the limit $h_0\to0$.

In order to relate $\Upsilon$ to the Helmholtz free energy 
$\Gamma_0$ we consider the Legendre transform
$\Gamma_0(r_0,u_0,M_0,k) =
F_0(r_0,u_0,h_0,k)  + M_0h_0$.
Differentiating $F_0=\Gamma_0-M_0h_0$ 
with respect to $k^2$ at fixed $h_0$ and making use of
$\partial\Gamma_0/\partial M_0=h_0$, we obtain
$(\partial F_0/\partial k^2)_{h_0} =
(\partial\Gamma_0/\partial k^2)_{M_0}$
and hence the bare helicity modulus
\begin{equation}
\Upsilon = \left. 2\frac{\partial}{\partial k^2}
\Gamma_0(r_0,u_0,M_0,k)\right|_{k=0} \label{eq:stiffness}
\end{equation}
where, in two-loop order, $\Gamma_0$ is given by 
Eq.~(\ref{eq:F_2loop}).
Thus, from Eq.~(\ref{eq:stiffness}), we obtain
\begin{eqnarray}
\Upsilon &=& 
M_0^2 + \Theta +2\lim_{k\rightarrow 0}\left(
\frac{\partial X_0}{\partial k^2}\right)_{\!\!M_0}
+O(u_0^2,u_0^2\ln u_0)\,, \label{eq:Ups_2} \\
\Theta &=& \int_{\bf p}\frac{1}{\bar{r}_{0L}+p^2}
+ \int_{\bf p}\frac{1}{\bar{r}_{0T}+p^2} -\frac{4}{k^2}
\int_{\bf p}
\frac{({\bf k\cdot p})^2}{(\bar{r}_{0L}+p^2)(\bar{r}_{0T}+p^2)}
\label{eq:UPS1}
\end{eqnarray}
where $\bar{r}_{0L}$ and $\bar{r}_{0T}$ are given by 
Eq.~(\ref{eq:r0L}) and 
the two-loop diagrams representing $X_0$ are given in Fig.~3.
The $d=3$ poles enter only through the order parameter $M_0^2$ 
since, being independent of 
\bf k\rm, they drop out of $X_0$ upon differentiation.
In $d=3$, therefore, $M_0^2$ is given for $h_0\to0$ by 
Eq.~(\ref{eq:M0_r0prime}) while in
$\Theta$ and $X_0$ we may simply replace $r_0$ by $r_0'$, 
the difference affecting only terms of
higher order. With this replacement, we write the one-loop 
contribution as $\Theta(r_0,u_0,M_0,k)=
\bare{\Theta}(r_0',u_0,M_0,k)+O(u_0^2,u_0^2\ln u_0)$; 
the corresponding two-loop contributions to the free
energy are denoted by $\bare{X}_A, \bare{X}_B,\ldots,\bare{X}_L$ 
(see Sec.~3 and App.~C).

Both $\bare{\Theta}$ and $2\partial \bare{X}/\partial k^2|_{k=0}$ 
contain Goldstone singularities
$\sim \bare{\chi}_{{}_T}^{1/2}\sim h_0^{-1/2}$ for $h_0\to0$ 
below $T_\c$. Expanding $\bare{\Theta}$ up to $O(u_0)$ we 
obtain for $r_0'<0$
\begin{eqnarray}
\bare{\Theta} &=& \left[
\frac{1}{12\pi}(-2r_0')^{1/2}+O(\alpha^{1/2})\, \right] +
\bare{\Theta}_1+O(u_0^2,u_0^2\ln u_0)\,, \label{eq:UPS1_2} \\
\bare{\Theta}_1 & = &\mbox{}-\frac{1}{8\pi^2}u_0\left[3\alpha^{-1/2}
+n-12+O(\alpha^{1/2}) \right]\,, \\ \label{eq:UPS12}
\alpha &=& \bare{\chi}_{{}_T}^{-1}(-2r_0'
+2\bare{\chi}_{{}_T}^{-1})^{-1} = w(1-w)^{-1}\,.
\label{eq:alpha0}
\end{eqnarray}
The other divergent terms
come from diagrams C and I in Fig.~\ref{fig:Gamma_0} and are given 
(see App.~C) by
\begin{eqnarray}
2\partial_{k^2} \bare{X}_C &=& 
\frac{1}{8\pi^2}u_0\left[\alpha^{-1/2}-\frac{8}{3}+O(\alpha^{1/2})
\right]\,,  \label{eq:g2c_3d_text} \\
2\partial_{k^2} \bare{X}_I &=&
\frac{1}{8\pi^2}u_0\left[2\alpha^{-1/2}-\frac{13}{3}
-\frac{16}{3}\ln2 +O(\alpha^{1/2})\,\right]\,,  
\label{eq:g2i_k_h_text}
\end{eqnarray}
These divergencies, however, are cancelled in the sum
\begin{equation}
\lim_{h_0\rightarrow 0} \left[ \bare{\Theta}_1 
+ 2\partial_{k^2} \bare{X}_C + 2\partial_{k^2} \bare{X}_I \right] =
\frac{1}{8\pi^2}u_0\left[5-n-\frac{16}{3}\ln2 \right]\,.
\label{eq:cancel}
\end{equation}
The remaining terms
are all finite for $h_0\rightarrow 0$ [see Appendix 
\ref{app:stiffness}] and their sum,
along with Eq.~(\ref{eq:cancel}), gives
\begin{eqnarray}
\Upsilon &=& \frac{(-2r_0')}{8u_0} + \frac{5}{6\pi}(-2r_0')^{1/2}
-\frac{1}{2\pi^2}(n+2)u_0 \ln \frac{(-2r_0')^{1/2}}{24u_0} 
\nonumber \\
&& -\, \frac{1}{6\pi^2} u_0\left[ n-15 -3(n-3)\ln3 \right]
+O(u_0^2,u_0^2\ln u_0)\,.  \label{eq:Ups_2loop}
\end{eqnarray}
for $h_0\to0$ and $r_0'<0$. Rewriting this in terms of the 
correlation length $\xi_-$ to 
absorb the logarithms of $u_0$ we obtain finally
\begin{eqnarray}
\Upsilon(\xi_-,u_0,3) =  \xi_-^{-1}\, \Bigg\{ && 
\,\frac{1}{8u_0\xi_-} + \frac{1}{24\pi} (3n+26) - \left[ 
\frac{1}{864\pi^2}(1169n-110) \right.\nonumber\\
&& - \left. \frac{1}{2\pi^2} (n-3)\ln3 \right] u_0\xi_- 
+ O(u_0^2\xi_-^2) \,\Bigg\}  \label{eq:Ups_xi-}
\end{eqnarray}
for the helicity modulus in three dimensions.
\subsection{Amplitude function of $\Upsilon$ in two-loop order}
The renormalization of the helicity modulus $\Upsilon(\xi_-,u_0,3)$ 
or superfluid density $\rho_s(\xi_-,u_0,3)$ requires only the 
renormalization of the coupling constant $u_0$. In renormalized 
form, $\Upsilon_R$ is obtained from $\Upsilon(\xi_-,u_0,d)$
according to
\begin{eqnarray}
\label{eq:ren_Upsilon}
\Upsilon_R(\xi_-,u,\mu,d) &=& \Upsilon(\xi_-,\mu^\epsilon 
Z_u Z_\phi^{-2} A_d^{-1} u, d) \\
&=& \xi_-^{2-d}A_{d}G(\mu\xi_-,u,d)\,.
\label{eq:Ups_general}
\end{eqnarray}
The quantity of interest here is the dimensionless amplitude 
function $G(1,u,d)$
which enters the final expression for $\Upsilon$ given below 
in Eq.~(\ref{eq:Ups_RGE}).
From Eqs.~(\ref{eq:Ups_xi-})--(\ref{eq:Ups_general}) with 
$Z_u(u,1)$ and $Z_\phi(u,1)$ 
in Eqs.~(\ref{eq:Z_u}) and (\ref{eq:Z_phi}) we obtain
\begin{equation}
G(1,u,3) = \frac{1}{8u}+\frac{1}{3} +
\left[ \frac{1}{54}(2378-683n)+8(n-3)\ln3 \right]u + O(u^2)
\label{eq:G_1u3}
\end{equation}
for the amplitude function in three dimensions. The first two 
terms agree with the previous \cite{schloms90,schloms87} 
one-loop result. The solution of the RG equation
\begin{equation}
\label{eq:RGG_Ups}
\left[ \partial_\mu + \beta_u \partial_u \right] 
\Upsilon_R (\xi_-,u,\mu,d) = 0\,,
\end{equation}
together with the choice $l_-\mu\xi_-=1$ for the flow
parameter, yields the following $(d=3)$ representation for 
the temperature dependence of $\Upsilon$ near $T_\c$
\begin{equation}
\Upsilon = (4\pi)^{-1} \xi_-^{-1} G(1,u(l_-),3) \label{eq:Ups_RGE}
\end{equation}
where the effective coupling $u(l_-)$ is determined by 
Eq.~(\ref{eq:RG_u_of_l}).
\subsection{Amplitude function of 
$\partial \bare{\chi}_{{}_T}(q)^{-1}/\partial q^2$ 
in two-loop order}
\label{subsec:chi_T}
\begin{figure}
\vspace*{-7mm}\hspace*{-22mm}\psfig{figure=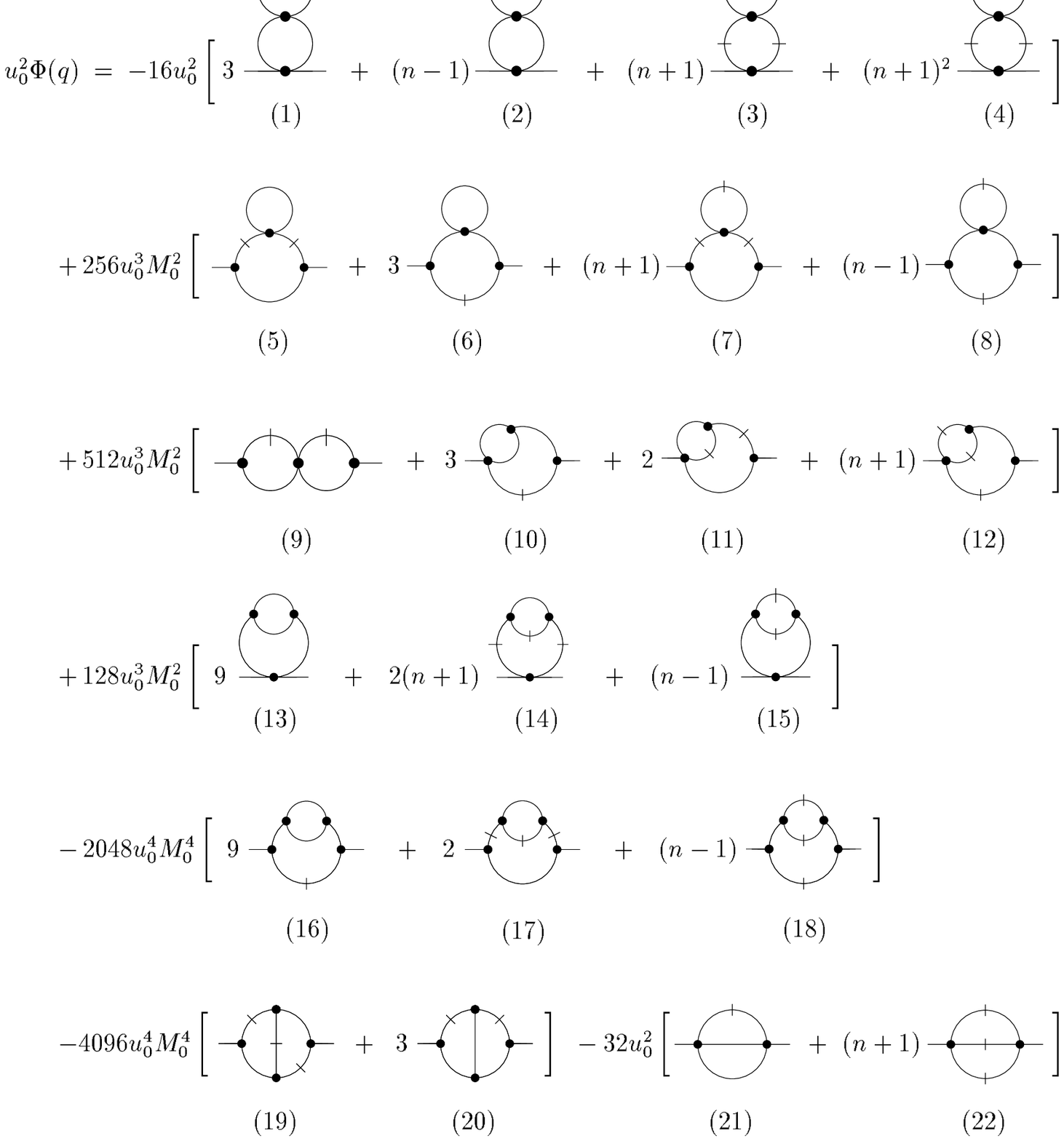,width=17cm}
\vspace*{-8cm}
\caption{Two-loop diagrams for the two-point vertex function
$\Gamma_{0T}^{(2)}(q)$, Eq.~(\protect\ref{eq:Gamma_T(q)}), 
with longitudinal and transverse
propagators $G_L=$ \protect\rule[2pt]{30pt}{0.1mm}\, and
$G_T=$ \protect\rule[2pt]{30pt}{0.1mm}\hspace{-15pt}
\protect\rule{0.1mm}{5pt}\hspace{17pt},
Eq.~(\protect\ref{eq:G_L}). The diagrams (5), (12), (17), (18),
(19) and (22) contain Goldstone singularities for $h_0\to0$ [see
Eqs.~(\protect\ref{eq:d_Gamma_T5})--(\protect\ref{eq:d_Gamma_T22})].}
\label{fig:Gamma_T}
\end{figure}

For $\rho_s$, there exists an alternative definition due to 
Josephson \cite{josephson66} in
terms of the transverse susceptibility $\bare{\chi}_{{}_T}(q)$ 
at finite wavenumber $q$
\begin{eqnarray}
\rho_s &=& k_BT \left( \frac{m}{\hbar} \right)^2 M_0^2\left. 
\frac{\partial}{\partial q^2}
\left[ \bare{\chi}_{{}_T}(q)^{-1} \right] \right|_{q=0}\,, 
\label{eq:josephson} \\
\bare{\chi}_{{}_T}(q) &=& \int\!\mbox{d}^{d}x\, 
\e^{\i{\bf q\cdot x}}
\langle\phi_{0i}({\bf x})\phi_{0i}(0)\rangle\,, 
\mbox{\hspace*{1cm}}i\neq 1  \label{eq:chiTq}
\end{eqnarray}
where the system is considered to be in a homogeneous state 
with $h_0\not=0$ (that is, with $k$
in Eqs.~(\ref{eq:h0_twist}) and (\ref{eq:M0_of_x}) set to zero). 
The inverse of $\bare{\chi}_{{}_T}(q)$ is equal to the 
transverse two-point vertex function
\begin{eqnarray}
\label{eq:Gamma_T(q)}
\Gamma_{0T}^{(2)}(q) &=&
\bar{r}_{0T} + q^2 + 4u_0 \int_{\bf p}\! G_L(p) + 4(n+1)u_0 
\int_{\bf p}\! G_T(p) \nonumber\\
&& -\, 64u_0^2M_0^2 \int_{\bf p}\! G_L(|{\bf p\!+\!q}|) G_T(p) 
+ u_0^2 \Phi(q) + O(u_0^3)
\end{eqnarray}
where $u_0^2\Phi(q)$ represents the negative sum of 22 two-loop 
digrams with two amputated transverse legs shown in 
Fig.~\ref{fig:Gamma_T}.
These diagrams are labelled by (1), (2), \ldots , (22).
The corresponding diagrams have also been given in Fig.~7 of 
Bervillier \cite{bervillier76} where,
however, our diagram (8) is missing and where the diagram 
corresponding to our diagram (4) has an incorrect prefactor. 
[These errors do not affect $\partial \Gamma_{0T}^{(2)} 
/\partial q^2|_{q=0}$, see Eq.~(\ref{eq:d_Gamma_T8}).] 
Bervillier \cite{bervillier76} calculated these diagrams within
the $\epsilon=4-d$ expansion.
Here we shall evaluate them in three dimensions.

We do not know of a rigorous proof of the equivalence between 
Josephson's definition of $\rho_s$, Eq.~(\ref{eq:josephson}),
and the helicity modulus defined in Eqs.~(\ref{eq:DF}) and 
(\ref{eq:rho_s}), but we shall
verify this equivalence here in two-loop order by calculating 
$\partial \bare{\chi}_{{}_T}(q)^{-1}/\partial q^2 |_{q=0}$ from 
Eq.~(\ref{eq:Gamma_T(q)}). The result
obtained for $\rho_s$ from Eq.~(\ref{eq:josephson}) provides an 
important independent check of our calculation of the 
helicity modulus. Since the calculation is nontrivial we also 
provide some of the intermediate results in Appendix D. 
In particular, we present in 
Eqs.~(\ref{eq:d_Gamma_T5})--(\ref{eq:d_Gamma_T22})
the Goldstone divergencies. They are more complicated than 
those of $\Upsilon$ but they are finally cancelled in the sum of 
the diagrammatic contributions. 
When expressed as a function of $r_0'$ we denote 
$\Gamma_{0T}^{(2)}(q)$ by $\bare{\Gamma}_T^{(2)}(q)$.
After a long calculation, requiring considerably more 
computational effort than the calculation
of $\Upsilon$ from the free energy, we obtain in three dimensions 
for $h_0\to0$ and $r_0'<0$
\begin{eqnarray}
\left.\frac{\partial}{\partial q^2}
\bare{\Gamma}_T^{(2)}(q)\right|_{q=0} &=& 1+
\frac{2}{3\pi}u_0(-2r_0')^{-1/2}+
\frac{1}{3\pi^2} (\,18-n-24\ln3\,) u_0^2(-2r_0')^{-1}
\nonumber \\  & & \mbox{}+O(u_0^{3},u_0^{3}\ln u_0)\,.
\label{eq:GT_k2_r0prime}
\end{eqnarray}
Eq.~(\ref{eq:GT_k2_r0prime}) contains no logarithms in $u_0$ 
since, at this order, there are no $d=3$ poles of 
$\partial\Gamma_{0T}^{(2)}(q)/\partial q^2|_{q=0}$.
In terms of the correlation length $\xi_-$ we obtain
\begin{eqnarray}
\left.\frac{\partial}{\partial q^2}
\bare{\Gamma}_T^{(2)}(\xi_-,u_0,q,3)\right|_{q=0} &=& 1+
\frac{2}{3\pi}u_0\xi_-+ \frac{2}{3\pi^2} (\,8-n-12\ln3\,) 
u_0^2\xi_-^2 \nonumber\\
&& +\, O(u_0^{3}\xi_-^{3})\,.  \label{eq:GT_k2_xi-}
\end{eqnarray}
Using Eqs.~(\ref{eq:josephson}) and (\ref{eq:M0_xi-}) this 
indeed reproduces $\Upsilon(\xi_-,u_0,3)$ as given in 
Eq.~(\ref{eq:Ups_xi-}) and thereby proves 
Eq.~(\ref{eq:rho_s}) up to two-loop order.

The renormalized vertex function $\Gamma_T^{(2)}$ is obtained 
from $\bare{\Gamma}_T^{(2)}$ according to
\begin{equation}
\label{eq:Ren_GammaT}
\Gamma_T^{(2)}(\xi_-,u,\mu,q,d) = Z_\phi 
\bare{\Gamma}_T^{(2)}(\xi_-,\mu^\epsilon A_d^{-1} Z_u 
Z_\phi^{-2} u,q,d)\,.
\end{equation}
On the basis of dimensional arguments we define the 
dimensionless amplitude function $f_T$ as \cite{schloms90}
\begin{equation}
\left.\frac{\partial}{\partial q^2} \Gamma_T^{(2)}(\xi_-,u,\mu,q,d)
\right|_{q=0} = f_T(\mu\xi_-,u,d)\,.
\label{eq:DGammaT_gen}
\end{equation}
Using Eqs.~(\ref{eq:Ren_GammaT}) and (\ref{eq:DGammaT_gen}) 
and substituting $Z_u(u,1)$ and $Z_\phi(u,1)$ we obtain 
in two-loop order in three dimensions
\begin{equation}
f_T(1,u,3) = 1+\frac{8}{3}u+ \left[ \frac{488}{3} -4n 
-128\ln3 \right]u^2+O(u^3)\,.  \label{eq:fT}
\end{equation}
The first two terms agree with the earlier one-loop result 
\cite{schloms90,schloms87}. Multiplying Eq.~(\ref{eq:fT}) by
the amplitude function $f_{\phi}$ in Eq.~(\ref{eq:f_phi2}) 
we indeed recover Eq.~(\ref{eq:G_1u3}) from 
Eqs.~(\ref{eq:rho_s}), (\ref{eq:Ups_RGE}) and 
(\ref{eq:josephson}) according to
\begin{equation}
\label{eq:agree}
G(1,u,3) = 4\pi f_\phi(1,u,3) f_T(1,u,3)\,.
\end{equation}

%% file: discussion.tex
%
%
\section{Results and discussion}
\label{sec:discussion}
\newcommand{\leqsim}{\,\raisebox{-1mm}{$\stackrel{{\textstyle < }}
                                           {{\textstyle \sim}}$}\,}
Within the minimally renormalized field theory in three dimensions 
we have derived 
the two-loop contributions to the amplitude
functions of the following $O(n)$ symmetric quantities of the 
$O(n)$ symmetric $\phi^4$ model
at vanishing external field:
\begin{itemize}
\item[(i)] the square of the order parameter 
$M_0^2=\langle\vec{\phi}_0\rangle^2$,
Eq.~(\ref{eq:f_phi2}),
\item[(ii)] the helicity modulus $\Upsilon$, Eq.~(\ref{eq:G_1u3}),
\item[(iii)] the $q^2$ part of the inverse of the transverse 
susceptibility 
$\mathaccent"7017{\chi}_{{}_T}(q)$,
Eq.~(\ref{eq:fT}), which yields the superfluid density $\rho_s$ 
(stiffness constant),
Eq.~(\ref{eq:josephson}),
\item[(iv)] the specific heat $\mathaccent"7017{C}^\pm$ above and 
below $T_\c$, 
Eqs.~(\ref{eq:F+2loop}) and (\ref{eq:F-2loop}).
\end{itemize}
Goldstone singularities arising in an intermediate stage of the 
calculations have been
shown to cancel among themselves. The resulting amplitude functions
are applicable to the asymptotic critical region as well as to the 
non-asymptotic region well away from criticality (apart from 
corrections arising from 
finite-cutoff effects, from $\phi^6$ terms and other higher-order 
couplings in $\cal H$, 
Eq.~(\ref{eq:LGW}), and from analytic terms). They provide the 
basis for
\begin{itemize}
\item[(a)] calculations of two-loop contributions to universal 
ratios $A_i/A_k$ and $a_i/a_k$
of leading and subleading amplitudes $A_i$ and $a_i$ which appear 
in the asymptotic representations \cite{privman91}
\begin{eqnarray}
\label{eq:phi_asymp}
M_0 &=& A_{{}_M} |t|^\beta \left( 1+a_{{}_M}|t|^\Delta + \ldots 
\right)\,, \\
\label{eq:rho_asymp}
\rho_s &=& A_{\rho_s} |t|^{(d-2)\nu} \left( 1 + 
a_{\rho_s}|t|^\Delta + \ldots \right)\,, \\
\label{eq:C_asymp}
\mathaccent"7017{C}^\pm &=& B + \frac{A^\pm}{\alpha} |t|^{-\alpha} 
\left( 1 + a_{{}_C}^\pm |t|^\Delta + \ldots \right)\,,
\end{eqnarray}
\item[(b)] nonlinear RG analyses of non-asymptotic critical 
phenomena of O($n$) symmetric
systems above and below $T_\c$ (for the general strategy of such 
analyses see 
Ref.~\cite{dohm87}).
\end{itemize}
These applications are of particular relevance to future 
experimental tests of the RG
predictions of critical-point universality along the $\lambda$ 
line of $^4$He $(n=2)$  
\cite{lipa95}.

\begin{figure}
\hspace*{-0.8cm}
\psfig{figure=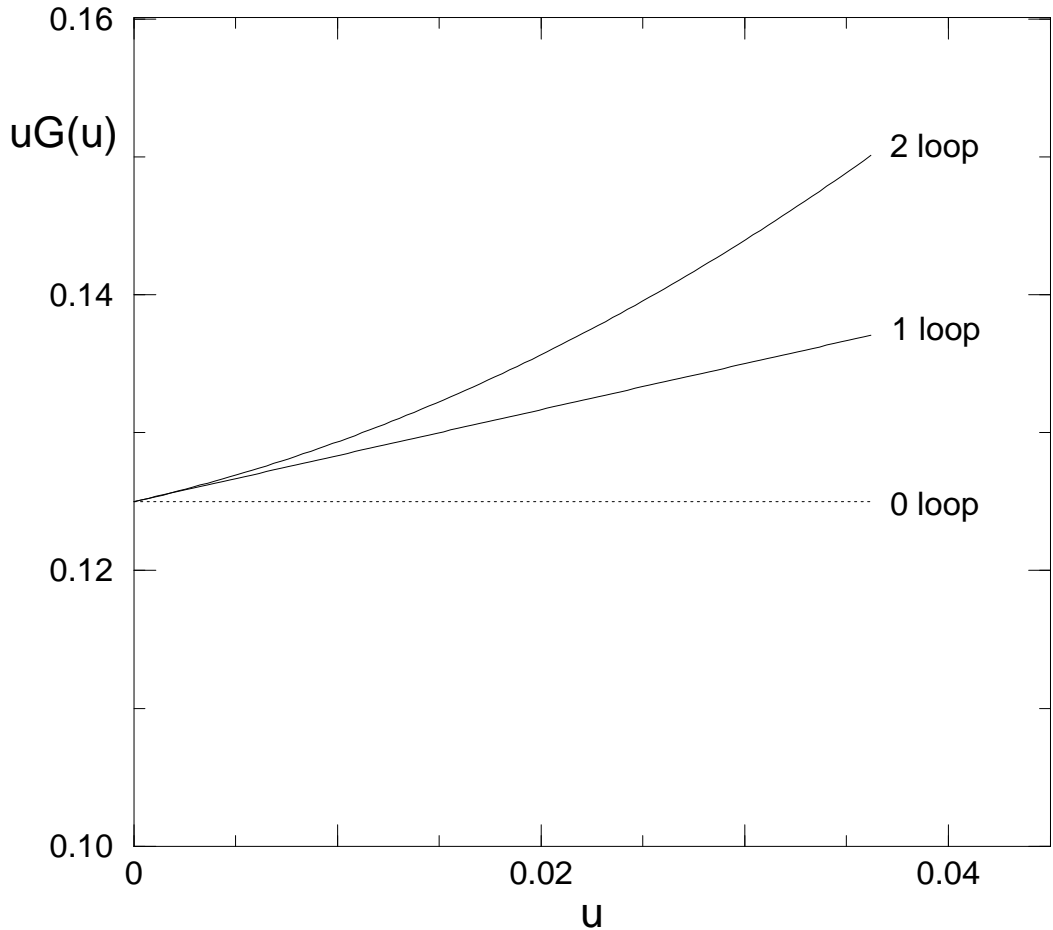,width=15.5cm}
\caption{Amplitude function $G(u)\equiv G(1,u,3)$ for the 
superfluid density
($n=2$), Eq.~(\ref{eq:G_1u3}), multiplied by $u$, as a function of 
the renormalized
coupling $u$ in zero-, one- and two-loop order.}
\label{fig:pl_G}
\end{figure}

\begin{figure}
\hspace*{-0.8cm}
\psfig{figure=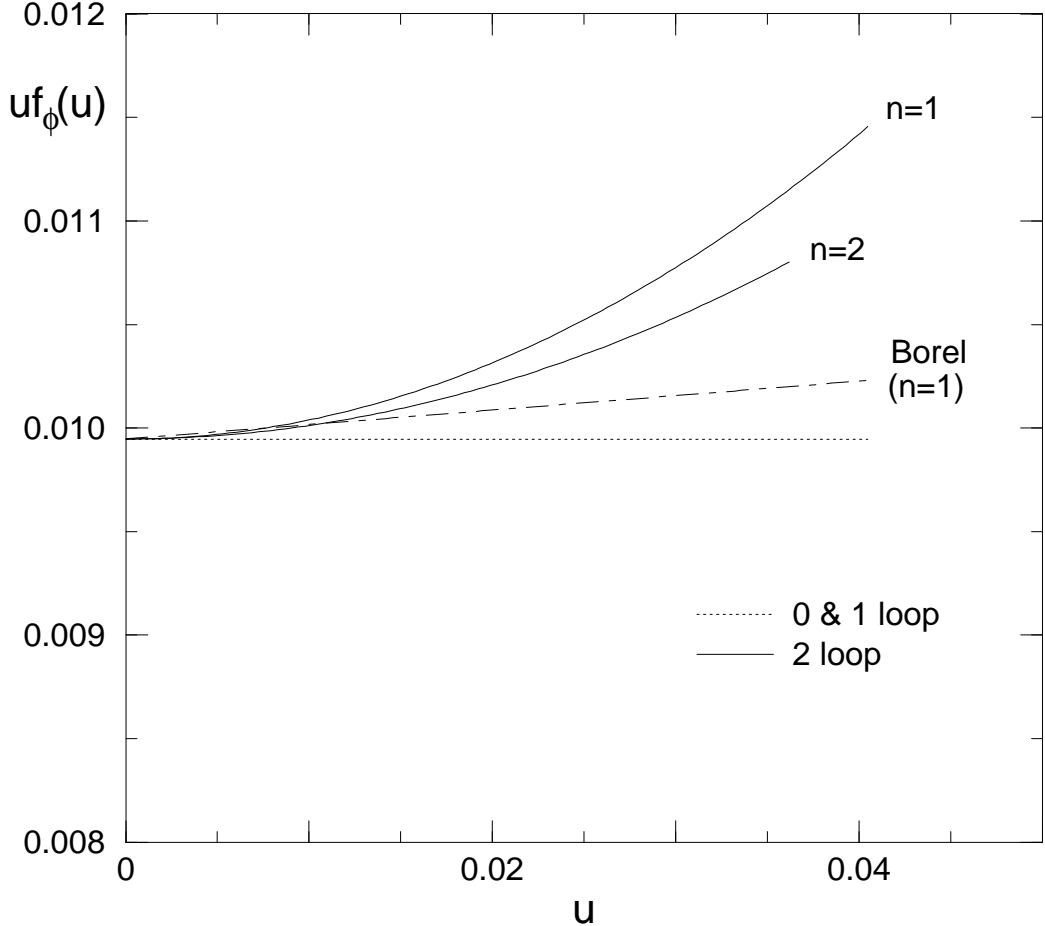,width=15.5cm}
\caption{Amplitude function $f_\phi(u)\equiv f_{\phi}(1,u,3)$ for 
the square of 
the order parameter ($n=1,\:2$), Eq.~(\ref{eq:f_phi2}), multiplied 
by $u$, as a function of the
renormalized coupling $u$ in zero-, one- and two-loop order. The 
dot-dashed
curve is the Borel summation result given for $n=1$ in 
Ref.~\protect\cite{halfkann92}.}
\label{fig:pl_fphi}
\end{figure}

\begin{figure}
\hspace*{-0.8cm}
\psfig{figure=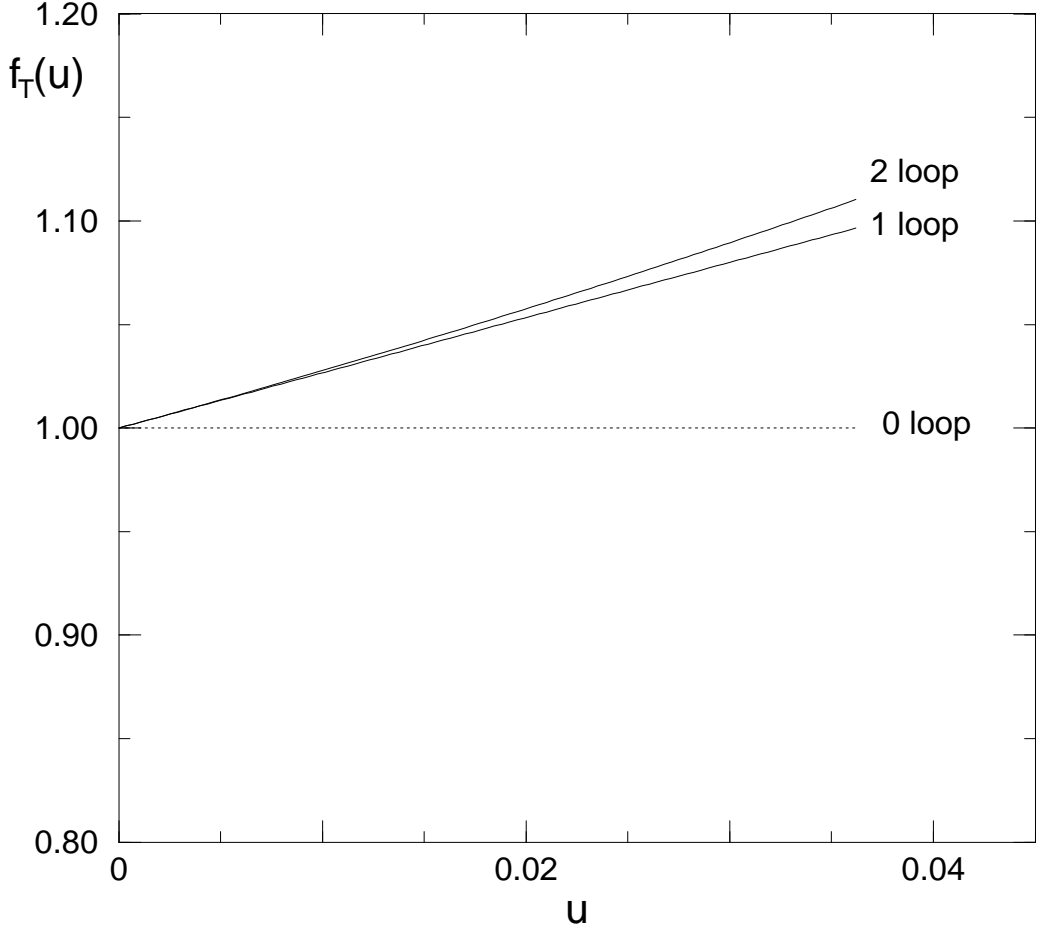,width=15.5cm}
\caption{Amplitude function $f_T(u)\equiv f_T(1,u,3)$ for
$\partial\mathaccent"7017{\chi}_{{}_T}(q)^{-1}/\partial q^2|_{q=0}$
for $n=2$, Eq.~(\ref{eq:fT}), as a function
of the renormalized coupling $u$ in zero-, one- and two-loop 
order.}
\label{fig:pl_fT}
\end{figure}

\begin{figure}
\hspace*{-0.8cm}
\psfig{figure=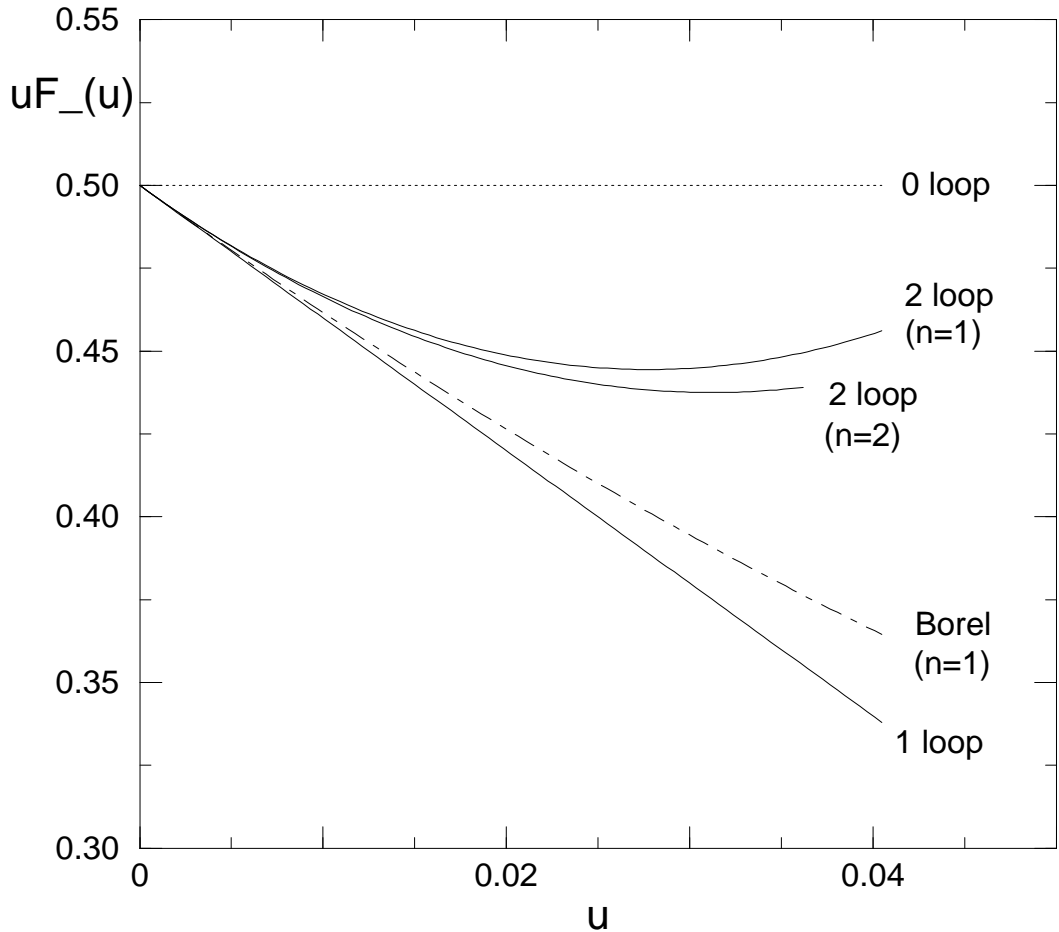,width=15.5cm}
\caption{Amplitude function $F_-(u)\equiv F_-(1,u,3)$ for the 
specific heat
below $T_\c$ ($n=$1, 2), Eq.~(\ref{eq:F-2loop}), multiplied by $u$,
as a function of the
renormalized coupling $u$ in zero-, one- and two-loop order. The 
dot-dashed
curve is the Borel summation result given for $n=1$ in 
Ref.~\protect\cite{halfkann92}.}
\label{fig:pl_F-}
\end{figure}

\begin{figure}
\hspace*{-0.8cm}
\psfig{figure=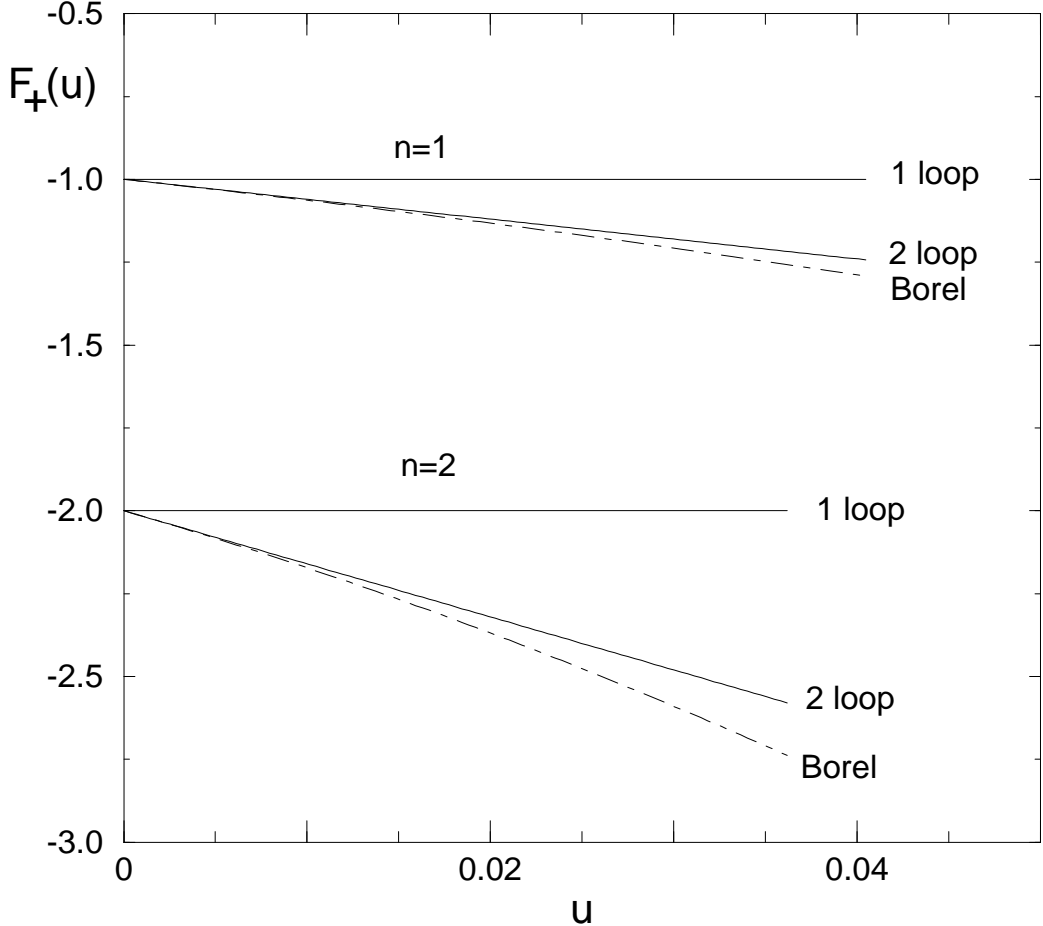,width=15.5cm}
\caption{Amplitude function $F_+(u)\equiv F_+(1,u,3)$ for the 
specific heat
above $T_\c$ ($n=$1, 2), Eq.~(\ref{eq:F+2loop}), as a function of 
the renormalized coupling $u$ in one- and two-loop order. The 
dot-dashed
curves are the Borel summation results given in 
Ref.~\protect\cite{krause90}.}
\label{fig:pl_F+}
\end{figure}

The various amplitude functions are plotted in 
Figs.~\ref{fig:pl_G}--\ref{fig:pl_F+}
vs the renormalized coupling
$u$ for the examples $n=1$ and $n=2$. In order to indicate the 
relative magnitude of the two-loop
contributions we have also plotted the corresponding zero- and 
one-loop approximations and,
where available, the Borel resummed results (for $n=1$ and $n=2$ 
above $T_\c$ \cite{krause90}
and for $n=1$ below $T_\c$ \cite{halfkann92}). The curves 
terminate at the fixed
points \cite{schloms89} $u^\star=0.0405$ for $n=1$ and 
$u^\star=0.0362$ for $n=2$
(although extensions to $u>u^\star$ may be needed in certain 
cases \cite{dohm93,bagnuls94,anisimov95}). We comment on these 
curves as follows.

For the superfluid density $(n=2)$, the one- and two-loop 
corrections for $G(1,u,3)$, see
Fig.~\ref{fig:pl_G}, 
each contribute about 10\% of the zero-loop term $1/8u^\star$ at 
the fixed point.
(For $n=3$ corresponding corrections are about 9\% and 5\%, 
respectively, at
$u^\star=0.0328$.) The fact that the one- and two-loop 
contributions are of comparable
magnitude suggests that higher-loop calculations including a 
Pad\'e-type analysis or a
Borel resummation are necessary before a reliable quantitative 
prediction can be made for
amplitude ratios such as \cite{schloms90} $R_\xi^T$ or 
$a_{{}_C}^-/a_{\rho_s}$. A preliminary
analysis \cite{burnett_unp} of experimental data 
\cite{ahlers72,tam85}
for $\rho_s$ and $\mathaccent"7017{C}^\pm$,
similar to the analyses in Refs.~\cite{dohm87,schloms87,dohm84}, 
indicate that the one-loop approximation for
$G(1,u^\star,3)$ is closer to the experimental result than the 
two-loop approximation.
Since $G(1,u,3)$ can be considered as consisting of two factors 
$4\pi f_\phi$ and $f_T$
according to Eq.~(\ref{eq:agree}), it is interesting to discuss 
the latter amplitude functions separately.

For the order parameter, we first consider $f_\phi$ for the case 
$n=1$ for which the Borel
summation result is known  \cite{halfkann92}. It is shown in 
Fig.~\ref{fig:pl_fphi} 
as the dot-dashed curve.
As pointed out previously \cite{halfkann92}, the Borel result 
deviates only very little
(by about 3\% at the fixed point) from the zero- and one-loop 
result whereas the two-loop
result is about 15\% larger at $u^\star$.
Obviously, the leading order approximation is the better one in 
this case.
We conjecture that this feature of the leading term
will remain true also when $n>1$ 
since the zero-loop term does not depend on $n$ and since the 
one-loop term vanishes for
general $n$ [due to the choice of the geometric factor $A_{d}$ in 
Eq.~(\ref{eq:ad})].
We note that for $n=2$ the two-loop result for $f_\phi$ shown in 
Fig.~6 lies about 9\%
above the zero- and one-loop results, similar to the two-loop term 
for the case $n=1$.
Our experience with other amplitude functions above $T_\c$ for 
which Borel
summation results are available \cite{krause90} is that the order 
in perturbation theory
which is closest to the Borel results seems to be the same for 
$n=1$, 2, 3.
Clearly, in order to make reliable quantitative predictions 
a higher-order calculation and Borel summation of the order 
parameter for $n>1$ are necessary.

Consider now the amplitude function $f_T$ of the transverse 
susceptibility
shown in Fig.~7. For $n=2$, 
the one-loop contribution at $u^\star$ is about 10\% whereas the 
two-loop correction
is much smaller, being only about 1\%. (For $n=3$ the corresponding
corrections are about 9\% and 1\%, respectively.)
We regard the smallness of the 
two-loop term as an indication of the quantitative reliability of 
the low-order formula 
for $f_T$
in Eq.~(\ref{eq:fT}). We infer from this and from our observations 
for $f_\phi$
that the considerable two-loop contribution to the amplitude 
function $G$ is mostly
due to the large two-loop term in $f_{\phi}$ and hence that the 
one-loop
approximation for $G$ is probably the most reliable at the present 
time.
A substantiation of this conjecture by higher-order calculations 
and Borel resummations
would be highly desirable. From a practical point of view, the 
presumed reliability of
the low-order result of $f_T$ is quite important since higher-order
calculations of $f_\phi$
for $n>1$ require considerably less computational effort than those
of $f_T$ or of $G$. Thus,
in a first step of future calculations of the amplitude function of
the superfluid density, 
it will be sufficient to perform higher-order
calculations only of $f_\phi$
for $n>1$ before embarking on a long-term project of much more 
difficult higher-order
calculations of $\mathaccent"7017{\chi}_{{}_T}(q)$ or of 
$\Upsilon$. 
We consider this important conclusion as a major result of our 
two-loop analysis.

These considerations give some support to the good agreement 
obtained
previously \cite{schloms87} between the one-loop formula for the 
superfluid
density and experimental data \cite{ahlers72,tam85} for $^{4}$He in
the 
(nonasymptotic) temperature range $10^{-6}\leqsim t\leqsim 10^{-2}$
at several pressures near the $\lambda$-line. 
A quantitative description of the nonasymptotic region, however, 
depends not only on amplitude functions like $G$ but also, 
crucially,
on an accurate knowledge of the effective coupling $u(l)$, which 
can be obtained
from the experimentally determined specific heat 
\cite{dohm85,dohm84} .
The formulas used to extract the nonuniversal initial value of the 
effective coupling
as a function of the pressure involve not only the RG exponent
functions, which are known accurately from
Borel summations \cite{schloms89}, but also the amplitude functions
of the specific heat
$F_{\pm}(1,u,3)$, which are given here to two-loop order in 
Eqs.~(\ref{eq:F+2loop}) and~(\ref{eq:F-2loop}).

Figure \ref{fig:pl_F-} shows the amplitude function $F_{-}$ of the
specific heat below $T_\c$
as a function of the renormalized 
coupling. As for the order parameter, we may use the known Borel 
summation
results \cite{halfkann92} for $n=1$ (dot-dashed curve in Fig.~7) to
infer the
reliability of the low order approximations for $n=2$. As noted 
previously \cite{halfkann92},
the one-loop approximation is the better one, differing from the 
Borel result at the fixed
point by about 5\% of the zero-loop term $1/2u^{\star}$ compared 
with about
16\% for the two-loop result. Evidently, the derivative of $F_-$ 
at $u=u^\star$,
which is needed
for universal correction-to-scaling ratios \cite{schloms90}, is 
also not well-approximated at 
two-loop order. Thus higher-order calculations of $F_-$ for $n>1$ 
are urgently needed.

In Fig.~\ref{fig:pl_F+}, we plot the amplitude function $F_{+}$ 
of the specific heat above $T_\c$,
where Borel results
are available for \mbox{$n\geq1$}, as a function of the 
renormalized
coupling \cite{krause90}. Here, 
it is the two-loop rather than the one-loop approximation which 
is closer
to the Borel results. This demonstrates that 
it is not clear {\em a priori\/} which (low) order of perturbation 
theory will provide the best approximation.

Let us note finally that the Borel results 
\cite{krause90,halfkann92} for
$F_{+}$ ($n\geq 1$) and for $F_{-}$ ($n=1$) in three dimensions 
neglect the
leading poles (in four dimensions) of the additive renormalization 
$A(u,\epsilon)$ in Eq.~(\ref{eq:A}) beyond two-loop order. Since 
a resummation
of the (as yet unknown) higher-order terms in the RG function 
$B(u)$ 
associated with $A(u,\epsilon)$ (see Eqs.~(2.32) and (2.33) in 
Ref.~\cite{schloms90}) 
is expected to yield a small correction \cite{nicoll85} of 
O($\eta$), we do 
not expect these Borel results for $F_\pm$ to be affected strongly 
by this 
approximation \cite{krause90}. However, at the level of accuracy 
anticipated in
future experiments \cite{lipa95}, it is likely that the present 
uncertainties regarding 
$B(u)$ will become significant, thus $B(u)$ itself will also be 
needed with the
improved accuracy provided by a resummation of its high-order
perturbation series. We recall that the function $B(u)$ enters not
only the formulas for the 
universal amplitude ratios \cite{schloms90,privman91} 
such as $A^+/A^-,\: R^T_\xi,\: a_{{}_C}^+/a_{{}_C}^-$ and
$a_{{}_C}^-/a_{\rho_s}$ but also the formulas needed to
determine the effective coupling $u(l)$ from the specific heat 
\cite{dohm85,dohm87,dohm84}.

In conclusion, our new two-loop results for $n>1$ below $T_\c$ 
provide additional motivation
and specific information on the strategy and direction of 
higher-order calculations planned
for future theoretical research parallel to the considerable effort
on the experimental side
to test the fundamental law of critical-point universality  
\cite{lipa95}.

%% file: appA.tex
%
%
\appendix
\section{Correlation lengths}
\label{app:corrl}
In this Appendix we derive Eqs.~(\ref{eq:r0prime+}) 
and~(\ref{eq:r0prime-}) and relate $\xi_\pm$ to $t$. 
Above $T_\c$, we obtain $\xi_+$ from Eq.~(\ref{eq:xi+2}) and the 
two-point vertex function \cite{schloms89} 
$\Gamma_0^{(2)}(q,r_0,u_0)=\bare{\chi}_+(q)^{-1}$. 
In two-loop order we have
\begin{eqnarray}
\lefteqn{ \Gamma_0^{(2)}(q,r_0,u_0) = r_0 +q^2 - 
\frac{4}{\epsilon}(n+2) A_d u_0 r_0^{1-\epsilon/2} 
+ \frac{16}{\epsilon^2}(n+2)^2 (1-\frac{\epsilon}{2})
A_d^2 u_0^2 r_0^{1-\epsilon} }\hspace{1cm} \nonumber\\
&-& 32(n+2) \frac{\Gamma(\epsilon)}{(4\pi)^d} \left[ 
\frac{J_0(\epsilon)}{\epsilon-1}
- J_2(\epsilon) q^2 +O(q^4) \right] u_0^2 r_0^{1-\epsilon} 
+ O(u_0^3)\,, \label{eq:Gamma^2(q)}\\
J_0(\epsilon) &=&
\int_0^{1}\mbox{d}x \int_0^{1}\mbox{d}y
\frac{y^{\epsilon/2 -1}}{[1-y+y(x-x^2)]^{2-\epsilon/2}}\,, 
\label{eq:J0} \\
J_2(\epsilon) &=& \int_0^{1}\mbox{d}x \int_0^{1}\mbox{d}y
\frac{y^{\epsilon/2}(1-y)(x-x^2)}{[1-y+y(x-x^2)]^{3-\epsilon/2}} 
\label{eq:J2}
\end{eqnarray}
which leads to
\begin{eqnarray}
\xi_+^{-2} &=& r_0 \left\{ 1-4(n+2) \frac{1}{\epsilon} 
A_d u_0 r_0^{-\epsilon/2}
+ 16(n+2)^2 \left[ \frac{1}{\epsilon^2} -\frac{1}{2\epsilon} \right] 
A_d^2 u_0^2 r_0^{-\epsilon} \right. \nonumber\\
&& \phantom{r_0 \Big[} - \left. a_2(\epsilon) u_0^2 r_0^{-\epsilon}
+O(u_0^3) \right\}\,,  \label{eq:xi+r0} \\
a_2(\epsilon) &=& 32(n+2) \Gamma(\epsilon) (4\pi)^{-d}
\left[ J_0(\epsilon)(\epsilon-1)^{-1}  + J_2(\epsilon) \right] 
\label{eq:a2}
\end{eqnarray}
where $J_0(1)=2\pi$ and $J_2(1)=2\pi/27$. 
Inversion of Eq.~(\ref{eq:xi+r0}) gives for $\xi_+^{-2}\not=0$
\begin{eqnarray}
r_0 &=& \xi_+^{-2} \left[ 1+4(n+2) \frac{1}{\epsilon} A_d u_0 
\xi_+^\epsilon + a_2(\epsilon) (u_0\xi_+^\epsilon)^2 
+O\! \left( (u_0\xi_+^\epsilon)^3 \right) \right]\,,
\label{eq:r0xi+}
\end{eqnarray}
where 
\begin{equation}
a_2(\epsilon) = \frac{n+2}{\pi^2} (\epsilon-1)^{-1} 
\left[ 1 + O(\epsilon-1) \right]  \label{eq:pole_r0c}
\end{equation}
determines the coefficient of the $d=3$ pole of $r_{0\c}$ in 
Eq.~(\ref{eq:r0C}).
Subtracting $\delta r_0=r_0-r_0'$ [see Eq.~(\ref{eq:shift})] 
from Eq.~(\ref{eq:r0xi+}) and letting $\epsilon\rightarrow 1$ 
we obtain
\begin{eqnarray}
r_0'&=& \xi_+^{-2}\, \Bigg\{\, 1 + \frac{1}{\pi}(n+2)u_0\xi_+
+ \frac{1}{27\pi^2} (n+2) (u_0\xi_+)^2 \left[ 1+ 54 \ln(24u_0\xi_+) 
\right] \nonumber \\
&& + \left[ 32(n+2)\frac{\partial}{\partial\epsilon}
\left.\frac{\Gamma(\epsilon)}{(4\pi)^{d}} J_0(\epsilon) 
\right|_{\epsilon=1}\!\!\!
-\frac{2}{\pi^2}(n+2)\ln 24 -C(n)\,\right](u_0\xi_+)^2 \nonumber\\
&& +\, O(u_0^3,u_0^3\ln u_0) \, \Bigg\}\, .  \label{eq:r0prime+C}
\end{eqnarray}
Here we have added and subtracted a logarithmic term 
$2\pi^{-2}(n+2)(u_0\xi_+)^2\ln24$
in order to conform\footnote[1]{See in particular Eq.~(A11) of 
Ref.~\cite{halfkann92}. Correspondingly
Eq.~(3.2) of Ref.~\cite{halfkann92} should read 
$\delta r_0=-3u_0^2\pi^{-2}(d-3)^{-1}+C(1)u_0^2$
with $C(1)$ given by our Eq.~(\ref{eq:C_eps}) for $n=1$. 
In line 18 after Eq.~(B5) of Ref.~\cite{halfkann92}, 
`` replacing $-(A_d')^b$ '' should read 
`` replacing $-(-A_d')^b$ ''.} 
with the representation of $r_0'$ in Refs.~\cite{bagnuls87} 
and \cite{halfkann92} for $n=1$ in terms of the bare coupling 
$g_0=24u_0$ of Ref.~\cite{bagnuls87}. 
Now a convenient choice for $C(n)$ is such that the term in the 
square brackets of Eq.~(\ref{eq:r0prime+C}) vanishes, i.~e.
\begin{equation}
C(n) = \frac{1}{\pi^2}(n+2) \left[ 1 - C_{\mathrm Euler} 
+ \ln \frac{4\pi}{9} -2\ln24 \right]\,.
\label{eq:C_eps}
\end{equation}
This yields Eq.~(\ref{eq:r0prime+}) and 
corresponds to the choice $C_1=0$ in Eq.~(A2) of 
Ref.~\cite{bagnuls87} for $n=1$.

Below $T_\c$, $\xi_-$ is defined,
according to Eqs.~(3.1)--(3.6) of Ref.~\cite{schloms90}, as
\begin{equation}
r_0'=\xi_-^{-2}R_{0+}(u_0\xi_-^{\epsilon},\epsilon)
-\frac{3}{2} \xi_-^{-2}Z_r(u(l_-),\epsilon)-\delta r_0\,, 
\label{eq:r0_xi-}
\end{equation}
(see also Eq.~(A4) in Ref.~\cite{halfkann92})
where the function $R_{0+}(u_0\xi_+^\epsilon,\epsilon)$ is 
defined above $T_\c$ and represents the series of $r_0$ in 
integer powers of $u_0\xi_+^\epsilon$
whose first three terms are given in the square brackets of 
Eq.~(\ref{eq:r0xi+}). The
effective coupling $u(l_-)$ as a function of the flow 
parameter $l_-$ is determined by Eq.~(\ref{eq:RG_u_of_l}) below.
For $\epsilon=1$, the function 
$\xi_-^{-2} R_{0+}(u_0\xi_-,1)-\delta r_0$ can be read off,
up to two-loop order,
from the right-hand side of Eq.~(\ref{eq:r0prime+}) which 
represents $\xi_+^{-2} R_{0+}(u_0\xi_+,1)-\delta r_0$. 
Thus it remains to rewrite $Z_r(u(l_-),1)$ as a function of 
$u_0\xi_-$ (see Eq.~(A7) of Ref.~\cite{halfkann92}). 
This leads to Eq.~(\ref{eq:r0prime-}).

The connection between $\xi_\pm$ and $t$ is given by 
\cite{schloms89,schloms90}
\begin{eqnarray}
  at &=& \xi_+^{-2}\, Q_+(1,u(l_+),d)\,
  \exp\int_{l_+}^1 \zeta_r(u(l))\frac{dl}{l}, \quad t>0,
\label{eq:t+} \\
-2at &=& \xi_-^{-2}\, Q_-(1,u(l_-),d)\,
  \exp\int_{l_-}^1 \zeta_r(u(l))\frac{dl}{l}, \quad t<0,
\label{eq:t-} \\
a &=& Z_r(u,\epsilon)^{-1}a_0\,, \label{eq:a}\\
\zeta_r(u) &=& \left. \mu \partial_\mu \ln Z_r(u,\epsilon)^{-1} 
\right|_0 \,,  \label{eq:RG_zeta_r_of_l} \\
l_\pm {\d}u(l_\pm)/{\d}l_\pm &=& \beta_u(u(l_\pm),\epsilon)\,, 
\label{eq:RG_u_of_l}
\end{eqnarray}
where 
$Q_\pm(1,u,d)$ are the amplitude functions of the correlation 
lengths $\xi_\pm$. They are related
according to $Q_-(1,u,d)=3-2Q_+(1,u,d)$. 
A convenient representation for $Q_+(1,u,3)$ is given in 
Eq.~(3.5) of Ref.~\cite{krause90}.

%% file: appB.tex
%
%
\section{Diagrams of $\Gamma_0^{(1)}$}
\label{app:eq_state}
In this appendix, we show the cancellation of (spurious) 
Goldstone divergences when the order parameter is calculated 
from the one-point vertex function 
$\Gamma_0^{(1)}(r_0,u_0,M_0)=h_0$
as given by the sum of 1PI diagrams with one amputated 
external leg. These diagrams have been given to two-loop order 
by Bervillier in Fig.~5 of Ref.~\cite{bervillier76}
and by Shpot in Eq.~(17) of Ref.~\cite{shpot90} where they were 
evaluated by use of an $\epsilon=4-d$ expansion. 
Here we shall work at $d=3$.
The analytic expression,up to order $O(u_0^2)$, is
\begin{eqnarray}
\frac{h_0}{M_0} = r_0+4u_0M_0^2
+12u_0 \int_{\bf p} G_L(p) +4(n-1)u_0 \int_{\bf p} G_T(p) 
+ u_0^2Y_0  \label{eq:M02_B}
\end{eqnarray}
where $Y_0=Y_{0a}+Y_{0b}+\ldots+Y_{0i}$ is the sum of the 
two-loop contributions
\begin{eqnarray}
\label{eq:Y_A}
Y_{0a} &=& -144 \int_{\bf p_1} G_L(p_1) \int_{\bf p_2} G_L(p_2)^2\,, 
\\  
\label{eq:Y_B}
Y_{0b} &=& -48(n-1) \int_{\bf p_1} G_T(p_1) \int_{\bf p_2} 
G_L(p_2)^2\,, \\
\label{eq:Y_C}
Y_{0c} &=& -16(n-1) \int_{\bf p_1} G_L(p_1) \int_{\bf p_2} 
G_T(p_2)^2\,, \\
\label{eq:Y_D}
Y_{0d} &=& -16(n^2-1) \int_{\bf p_1} G_T(p_1) \int_{\bf p_2} 
G_T(p_2)^2\,, \\
\label{eq:Y_E}
Y_{0e} &=& -96 \int_{\bf p_1} \int_{\bf p_2} G_L(p_1)\, G_L(p_2)\, 
G_L(|\bf p_1\!+\!p_2|)\,, \\
\label{eq:Y_F}
Y_{0f} &=& -32(n-1) \int_{\bf p_1} \int_{\bf p_2} G_T(p_1)\, 
G_T(p_2)\,G_L(|\bf p_1\!+\!p_2|)\,, \\ 
\label{eq:Y_G}
Y_{0g} &=& 3456u_0M_0^2
\int_{\bf p_1} \int_{\bf p_2} G_L(p_1)^2\, G_L(p_2)\, 
G_L(|\bf p_1\!+\!p_2|)\,, \\
\label{eq:Y_H}
Y_{0h} &=& 384(n-1)u_0M_0^2
\int_{\bf p_1} \int_{\bf p_2} G_T(p_1)\, G_T(p_2)\, 
G_L(|{\bf p_1\!+\!p_2}|)^2\,, \\
\label{eq:Y_I}
Y_{0i} &=& 256(n-1)u_0M_0^2
\int_{\bf p_1} \int_{\bf p_2} G_T(p_1)^2\, G_T(p_2)\, 
G_L(|\bf p_1\!+\!p_2|)
\end{eqnarray}
with $G_L(p)$ and $G_T(p)$ given by Eq.~(\ref{eq:G_L}).
Eqs.~(\ref{eq:M02_B})--(\ref{eq:Y_H}) agree with 
Refs.~\cite{shpot90} and \cite{bervillier76}.
The prefactor $256(n-1)u_0M_0^2$ in Eq.~(\ref{eq:Y_I}) agrees 
with the prefactor of the corresponding diagram of 
Shpot \cite{shpot90} who corrected the corresponding prefactor of
Bervillier \cite{bervillier76}.

The quantities $Y_{0e}$ and $Y_{0f}$ contain $d=3$ poles, but since
$Y_{0e}=2(u_0^2M_0^2)^{-1}X_{0H}$ and
$Y_{0f}=2(u_0^2M_0^2)^{-1}(X_{0I}+X_{0J})$ for $k=0$
[see Eqs.~(\ref{eq:g2h})--(\ref{eq:g2j})],
the substitution $r_0=r_0'+\delta r_0$ in Eq.~(\ref{eq:M02_B})
leads to the cancellation of these poles in the same way as in 
Eq.~(\ref{eq:f}). Thus,
\begin{equation}
\label{eq:cancellation}
\lim_{\epsilon\to 1}\,
\left[ \delta r_0 + u_0^2(\bare{Y}_e + \bare{Y}_f) \right] 
= 2u_0^2 f(r_0',u_0,M_0)
\end{equation}
with $f(r_0',u_0,M_0)$ given by Eq.~(\ref{eq:f_r0}).
At this stage, one has the choice of performing the integrations
before or after solving iteratively for $M_0^2$.
Carrying out the integrations first, one is led to 
Eq.~(\ref{eq:h0tilde}). Inverting first, one finds
\begin{eqnarray}
M_0^2 &=& \frac{1}{4u_0} (-r_0'+\bare{\chi}_{{}_T}^{-1})
+\frac{3}{4\pi} (-2r_0'+3\bare{\chi}_{{}_T}^{-1})^{1/2}
+\frac{1}{4\pi}(n-1) \bare{\chi}_{{}_T}^{-1/2} + u_0 \bare{Y}_1 
\nonumber\\
&& -\, \frac{u_0}{4} \left[ \bare{Y} -\bare{Y}_{e} -\bare{Y}_{f} 
\right] - \frac{u_0}{2} \tilde{f}(r_0',u_0,\bare{\chi}_{{}_T})
+O(u_0^2,u_0^2\ln u_0) \label{eq:M02_C}
\end{eqnarray}
where
\begin{equation}
\label{eq:tilde_f}
\tilde{f}(r_0',u_0,\bare{\chi}_{{}_T}) = \frac{n+2}{\pi^2} \ln 
\frac{(-2r_0'+3\bare{\chi}_{{}_T}^{-1})^{1/2}}{24u_0} 
+ \frac{n-1}{\pi^2} \ln \frac{1+2w^{1/2}}{3} \,,
\end{equation}
with $w$ given by Eq.~(\ref{eq:w,lambda}), that is, 
$\tilde{f}=f$ in Eq.~(\ref{eq:f_r0}) in lowest order in $u_0$, and
\begin{eqnarray}
u_0\bare{Y}_1 &=& \frac{u_0}{8\pi^2} \left\{ 27 + (n-1) \left[ 
n-1 +9w^{1/2} +3w^{-1/2} \right] \right\} \label{eq:Y1} 
\end{eqnarray}
is the $O(u_0)$ contribution from the expansion of the integrals 
in Eq.~(\ref{eq:M02_B}) at one-loop order.
In three dimensions we obtain for finite $w$
\begin{eqnarray}
\bare{Y}_a&=& 9\, (2\pi^2)^{-1}\,, \label{eq:YA} \\
\bare{Y}_b&=& 3\, (2\pi^2)^{-1} (n-1) w^{1/2}\,, \label{eq:YB}\\
\bare{Y}_c&=&   (2\pi^2)^{-1} (n-1) w^{-1/2}\,, \label{eq:YC}\\
\bare{Y}_d&=&   (2\pi^2)^{-1} (n^2-1)\,, \label{eq:YD} \\
\bare{Y}_g&=& 9\, (2\pi^2)^{-1} (1-w)\,, \label{eq:YG} \\
\bare{Y}_h&=& 3\, (2\pi^2)^{-1} (n-1)(1-2w^{1/2})(1-w)(1-4w)^{-1}\,,
\label{eq:YH} \\
\bare{Y}_i&=& \pi^{-2} (n-1) (w^{-1/2}-2)(1-w)(1-4w)^{-1}\,. 
\label{eq:YI}
\end{eqnarray}
Goldstone divergences $\sim w^{-1/2}$ arise from 
Eqs.~(\ref{eq:Y1}), (\ref{eq:YC}), and (\ref{eq:YI}) 
for $h_0\to0$. They cancel among themselves
in $\bare{Y}_1-\quart(\bare{Y}_c+\bare{Y}_i)$.
Summing the remaining terms, one obtains the same result, 
Eq.~(\ref{eq:M02_of_h}), as derived via the free energy 
$\bare{\Gamma}$.

%% file: appC.tex
%
%
\section{Two-loop diagrams of $\Gamma_0$ at $\bf k\not=0$}
\label{app:stiffness}
The contributions $X_{0A}$, $X_{0B}$, \ldots,
$X_{0L}$ in Eq.~(\ref{eq:F_2loop}) at finite $k$
are represented by the two-loop diagrams in Fig.~3.
Diagrams A--G consist of products of exactly calculable one-loop 
integrals \cite{haussmann92}.
The integral expressions of the diagrams H--L read
\begin{eqnarray}
X_{0H}&=&-48u_0^{2}M_{0}^{2}
\int_{\bf p_1} \int_{\bf p_2}
R_L(p_1)\, R_L(p_2)\, R_L(|{\bf p_1+p_2}|)\,, \label{eq:g2h} \\
X_{0I}&=&-16u_0^{2}M_{0}^{2} 
\int_{\bf p_1} \int_{\bf p_2} R_T(p_1)\, R_T(p_2)\, 
R_L(|{\bf p_1+p_2}|)\,, 
\label{eq:g2i} \\
X_{0J}&=&-16(n-2)u_0^{2}M_{0}^{2}\int_{\bf p_1}
\int_{\bf p_2} G_T(p_1)\, G_T(p_2)\, R_L(|{\bf p_1+p_2}|)\,, 
\label{eq:g2j} \\
X_{0K}&=&-96u_0^{2}M_{0}^{2} 
\int_{\bf p_1} \int_{\bf p_2} 
R_{LT}(p_1)\, R_{LT}(p_2)\, R_L(|{\bf p_1+p_2}|)\,, \label{eq:g2k}\\
X_{0L}&=&-32u_0^{2}M_{0}^{2}
\int_{\bf p_1} \int_{\bf p_2}
R_{LT}(p_1)\, R_{TL}(p_2)\, R_T(|{\bf p_1+p_2}|)\,.
\label{eq:g2l} 
\end{eqnarray}
Here we have used the elements of the matrix $K^{-1}$ 
of $\mathbf k$-dependent propagators
\begin{eqnarray}
K({\mathbf p,k})^{-1} &=& \left(\begin{array}{cc} R_L(p)&R_{LT}(p)\\
R_{TL}(p)&R_T(p) \end{array} \right)\,, \label{eq:matrix}\\
R_L(p) &=& \frac{\bar{r}_{0T}+p^2+k^2}{\det K({\bf p,k})}\,, \qquad
R_{LT}(p) = \frac{-2i\,\bf k\!\cdot\! p}{\det K({\bf p,k})}\,, 
\label{eq:R_L} \\
R_T(p) &=& \frac{\bar{r}_{0L}+p^2+k^2}{\det K({\bf p,k})}\,, 
\qquad R_{TL}(p) = -R_{LT}(p) 
\label{eq:R_LT}
\end{eqnarray}
where $K(\bf p,k)$, $\bar{r}_{0L}$, $\bar{r}_{0T}$, $G_L(p)$ 
and $G_T(p)$ are given in Eqs.~(\ref{eq:matrix_K})--(\ref{eq:G_L}), 
respectively.
The contributions $\partial\bare{X}_{j}/\partial k^2|_{k=0}$
to the helicity modulus, Eq.~(\ref{eq:Ups_2}), will be denoted by 
$\partial_{k^2}\bare{X}_j$.
The quantities 
$\partial_{k^2}\bare{X}_C=\partial_{k^2}[2u_0
\int_{\bf p_1}R_L(p_1)\int_{\bf p_2} 
R_T(p_2)]$ and $\partial_{k^2}\bare{X}_I$
are divergent for $h_0\to0$. While $\partial_{k^2}\bare{X}_C$ 
is easily evaluated in terms of standard one-loop integrals, 
$\partial_{k^2}\bare{X}_I$ is given, in three dimensions, by
\begin{eqnarray}
\label{eq:II}
2 \partial_{k^2} \bare{X}_I &=&
\frac{-4/3}{(4\pi)^3}\, u_0 \Big[\, 8(1+\alpha) I_1  +(1+4\alpha) 
I_2 +4\alpha I_3 - ( 6+8\alpha ) I_4 \,\Big]\,, \\
I_i &=& \int_0^1{\d}x \int_0^1 {\d}y \:
\frac{y^{-1/2}\, f_i(x,y)}{[1-y+y(x-x^{2})]^{3/2}}\,, 
\label{eq:I_i}\\
f_1(x,y) &=& \ln (y+\alpha) - \ln(1-y+\alpha)\,, \quad
f_2(x,y)  =  xy\,(xy+\alpha)^{-1}\,, \label{eq:f_1,2}\\
f_3(x,y) &=& \ln\alpha - \ln(1-y+\alpha)\,, \quad
f_4(x,y)  =  xy\,(1-y+\alpha)^{-1}\,, \label{eq:f_3,4}
\end{eqnarray}
where $\alpha$ is given by Eq.~(\ref{eq:alpha0}). 
To illustrate the evaluation of these integrals, we consider 
$I_4$. Thus,
\begin{eqnarray}
I_4 &=& \int_0^{1}{\d}x \int_0^{1}{\d}y
\frac{y^{1/2}x\, (1-y+\alpha)^{-1}}{[1-y+y(x-x^2)]^{3/2}} 
\label{eq:g1}\\
&=& \int_{1/4}^\infty {\d}z \int_0^{1/2}{\d}u
\frac{(z-u^2)^{3/2}}{[z-1/4+\alpha(z+3/4)]} \label{eq:g2}\\
&=& \int_{1/4}^\infty \frac{{\d}z}{z}\, 
\frac{(4z-1)^{-1/2}}{[z-1/4+\alpha(z+3/4)]} \label{eq:g3}\\
&=& \frac{8}{1+\alpha} \int_0^\infty {\d}t \,\,
\frac{1}{(t^2+1)}\,\frac{1}{(t^2+4\alpha(1+\alpha)^{-1})} 
\label{eq:g4} \\
&=& \frac{2\pi}{1-3\alpha} \left[ \alpha^{-1/2}(1+\alpha)^{1/2} 
-2 \right]\,. \label{eq:g5}
\end{eqnarray}
In going from (\ref{eq:g1}) to (\ref{eq:g5}) we have used the 
substitutions $z=y^{-1}-3/4$,
$u=x-1/2$, then $u=z^{1/2}\sin w$ and finally $z=(t^2+1)/4$.
While $I_4$ exhibits a divergence for $h_0\to0$, the
other contributions to $\partial_{k^2}\bare{X}_I$ are finite in
this limit and lead to Eq.~(\ref{eq:g2i_k_h_text}).

The remaining contributions are finite for $h_0\to0$. 
For $j=$ A, B and F they are readily 
evaluated in three dimensions as
{
\mathindent10pt
\begin{eqnarray} 
2 \partial_{k^2} \bare{X}_{A} 
&=& \textstyle 2 \partial_{k^2} \Big[ 3u_0 \left(\int_{\bf p}
R_L(p)\right)^2 \Big] = -(8\pi^2)^{-1} u_0\, , \label{eq:IA} \\ 
2 \partial_{k^2} \bare{X}_{B}
&=& \textstyle 2 \partial_{k^2} \Big[ 3u_0 \left(\int_{\bf p}
R_T(p)\right)^2 \Big]= 3\, (8\pi^2)^{-1} u_0\, ,  \label{eq:IB} \\ 
2 \partial_{k^2} \bare{X}_{F}
&=& \textstyle 2 \partial_{k^2} \Big[ 2(n-2)u_0 \int_{\bf p_1} \!
R_T(p_1)\! \int_{\bf p_2}\! G_T(p_2) \Big] = 
(8\pi^2)^{-1} (n-2)u_0\,.  \label{eq:IF}
\end{eqnarray}
}
$\partial_{k^2}\bare{X}_E=\partial_{k^2}[2(n-2)u_0
\int_{\bf p_1}R_L(p_1)\int_{\bf p_2} G_T(p_2)]$ 
vanishes for $h_0\to0$, 
$\bare{X}_G=n(n-2)u_0\,[\int_{\bf p}G_T(p)]^2$ is $k$-independent, 
and $\bare{X}_D=4u_0\,[\int_{\bf p}R_{LT}(p)]^2=0$.
For $j=$ H, J, K and L we use splitting by partial fraction 
and obtain in three dimensions for $h_0\to0$
\begin{eqnarray}
2 \partial_{k^2} \bare{X}_H
&=& -(8\pi^2)^{-1} u_0 \, ,  \label{eq:IH} \\ 
2 \partial_{k^2} \bare{X}_J
&=& -(24\pi^2)^{-1} (n-2)u_0 \,, \label{eq:IJ} \\
2 \partial_{k^2} \bare{X}_K &=& \pi^{-2} (1-\ln3)u_0\,, 
\label{eq:IK} \\
2 \partial_{k^2} \bare{X}_L &=& (3\pi^2)^{-1} (2\ln2-1)u_0\,.  
\label{eq:IL}
\end{eqnarray}

%% file: appD.tex
%
%
\section{Contributions to $\partial\Gamma_{0T}^{(2)}/\partial q^2$}
\label{app:chi_T}
We denote the integral expressions of the two-loop contribution 
$\Phi(q)$ in Eq.~(\ref{eq:Gamma_T(q)}) by 
$\Phi_i$, $i=1$, 2, \ldots , 22.
The derivative of $\Gamma_{0T}^{(2)}$ with respect to $q^2$ at 
$q=0$ yields
\begin{eqnarray}
\label{eq:dGamma_dq^2}
\left. \partial_{q^2} \Gamma_{0T}^{(2)}(q) \right|_{q=0}
&=& 1 + u_0 \Psi +u_0^2 \left. \partial_{q^2}\, \Phi(q) 
\right|_{q=0} + O(u_0^3)\,, \\
\label{eq:Psi}
\Psi &=& \frac{8}{d}\cdot 
\frac{\epsilon\bar{r}_{0T}+d\bar{r}_{0L}}{(\bar{r}_{0L}
-\bar{r}_{0T})^2} \int_{\bf p} \left[ G_T(p)-G_L(p) 
-(\bar{r}_{0L}-\bar{r}_{0T}) G_L(p)^2 \right] \nonumber\\
&& -\, \frac{32}{d} \bar{r}_{0L} \int_{\bf p}\! G_L(p)^3\,.
\end{eqnarray}
Expanding $\Psi$ with respect to $u_0$ and replacing 
$r_0$ by $r_0'$ we obtain in three dimensions
\begin{eqnarray}
\label{eq:Exp_Psi}
u_0\Psi &=& \frac{2}{3\pi}u_0 \tau \left[ 1+ O(\alpha^{1/2}) 
\right] +u_0^2 \Psi_1 + O(u_0^3)\,, \\
\label{eq:Psi_1}
\Psi_1 &=& -\frac{1}{\pi^2} \tau^2 \left[ 3\alpha^{-1/2} 
+n-8 +O(\alpha^{1/2}) \right]\,, \\
\tau &=& (-2r_0'+2\bare{\chi}_{{}_T}^{-1})^{-1/2}\,. \label{eq:tau}
\end{eqnarray} 
Eq.~(\ref{eq:Psi_1}) exhibits a Goldstone singularity 
$\sim\alpha^{-1/2}$ with $\alpha$ given by Eq.~(\ref{eq:alpha0}).
Further Goldstone singularities are contained in the 
two-loop diagrams (5), (12), (17), (18), (19) and (22) 
in Fig.~\ref{fig:Gamma_T}. We find, in three dimensions,
\begin{eqnarray}
\label{eq:d_Gamma_T5}
\partial_{q^2} \Phi_5
&=& \pi^{-2}  \tau^2 \left[ \alpha^{-1/2}
-4 +O( \alpha^{1/2} ) \right], \\
\partial_{q^2} \Phi_{12}
&=& -2\, (27\pi^2)^{-1} (n+1)  \tau^2 \left[
\alpha^{-1} -9\ln(9\alpha) -12 +O( \alpha^{1/2} ) \right], 
\label{eq:d_Gamma_T12}\\
\label{eq:d_Gamma_T17}
\partial_{q^2} \Phi_{17}
&=& (6\pi^2)^{-1}  \tau^2 \left[ 12\alpha^{-1/2} 
-7-80\ln2+O( \alpha^{1/2} ) \right], \\
\partial_{q^2} \Phi_{18}
&=& (27\pi^2)^{-1}(n-1)  \tau^2 \left[
\alpha^{-1} -18\ln(9\alpha) -33 +O( \alpha^{1/2} ) \right], 
\label{eq:d_Gamma_T18}\\
\label{eq:d_Gamma_T19}
\partial_{q^2} \Phi_{19}
&=& (54\pi^2)^{-1}  \tau^2 \left[ 4\alpha^{-1} 
-72\ln(36\alpha) -87 +O( \alpha^{1/2} ) \right]\,, \\
%
%
\label{eq:d_Gamma_T22}
\partial_{q^2} \Phi_{22} &=& (27\pi^2)^{-1}(n+1) \tau^2 
\alpha^{-1} \,,
\end{eqnarray}
where $\partial_{q^2} \Phi_i=\partial \Phi_i/\partial q^2|_{q=0}$. 
Summing up these contributions we get the finite result for 
$h_0\to0$
\begin{eqnarray}
\lefteqn{ \lim_{h_0\to0}\, \left\{ \Psi_1 + \partial_{q^2} 
\left[ \Phi_5 + \Phi_{12} + \Phi_{17} + \Phi_{18} + \Phi_{19} 
+\Phi_{22} \right] \right\} }\hspace{1cm}
\nonumber\\
&=& (3\pi^2)^{-1} (-2r_0')^{-1} \left[ 10 - 4n -48\ln 2 \right]\,.
\label{eq:finite}
\end{eqnarray}
The $q$-independent diagrams (1)--(4) and (13)--(15) do not 
contribute. The remaining diagrams give for $h_0\to0$ in 
three dimensions
\begin{eqnarray}
\label{eq:d_Gamma_T6}
\partial_{q^2} \Phi_6
&=& 3\pi^{-2}  (-2r_0')^{-1} \,, \\
\label{eq:d_Gamma_T7}
\partial_{q^2} \Phi_7
&=& (n+1)\pi^{-2} (-2r_0')^{-1} \,, \\
\label{eq:d_Gamma_T8}
\partial_{q^2} \Phi_8
&=& 0 \,, \\
\label{eq:d_Gamma_T9}
\partial_{q^2} \Phi_9
&=& -8\, (3\pi^2)^{-1}  (-2r_0')^{-1} \,, \\
\label{eq:d_Gamma_T10}
\partial_{q^2} \Phi_{10}
&=& -(2\pi^2)^{-1}  (-2r_0')^{-1} \left[ 1+8\ln (3/2) \right] \,, \\
\label{eq:d_Gamma_T11}
\partial_{q^2} \Phi_{11}
&=& \pi^{-2}  (-2r_0')^{-1} \left[ 3-8\ln2 \right] \,, \\
\label{eq:d_Gamma_T16}
\partial_{q^2} \Phi_{16}
&=& (2\pi)^{-2}  (-2r_0')^{-1} \left[ 48\ln (3/2) -1 \right] \,, \\
\label{eq:d_Gamma_T20}
\partial_{q^2} \Phi_{20}
&=& -\pi^{-2}  (-2r_0')^{-1} \left[ 1 +16\ln(3/4) \right] \,, \\
\label{eq:d_Gamma_T21}
\partial_{q^2} \Phi_{21}
&=& (12\pi^2)^{-1}  (-2r_0')^{-1} \,.
\end{eqnarray}
This leads to Eq.~(\ref{eq:GT_k2_r0prime}).

%% file: refs.tex
%
%